\documentclass[aps, prb, showpacs, reprint, preprintnumbers,amsmath,amssymb,superscriptaddress,floatfix]{revtex4-2}

\usepackage[]{graphicx}      
\usepackage{bm}             	
\usepackage{times}			
\usepackage{tabularx}
\usepackage{color}
\usepackage{dcolumn}		
\usepackage{amsmath}
\usepackage{amssymb}
\usepackage{amsfonts}
\usepackage{calc}
\usepackage{times}
\usepackage{tabularx}
\usepackage{xcolor,colortbl}
\usepackage{rotating}
\usepackage{hyperref}
\usepackage{amsmath}

\makeatletter
\newcommand\xleftrightarrow[2][]{%
  \ext@arrow 9999{\longleftrightarrowfill@}{#1}{#2}}
\newcommand\longleftrightarrowfill@{%
  \arrowfill@\leftarrow\relbar\rightarrow}
\makeatother

\newcommand*{\xdash}[1][3em]{\rule[0.5ex]{#1}{0.55pt}}
\newcommand{\sZ}{\ensuremath{\scalebox{.86}{Z}
\kern-.41em\raisebox{.07em}{$\xdash[.3em]$}}}
\newcommand{\sz}{\ensuremath{\scalebox{.86}{z}
\kern-.3em\raisebox{-.045em}{$\xdash[.2em]$}}}

\begin{document}
\title{Theory of propagating spin wave spectroscopy using inductive antennas: conditions for unidirectional energy flow}
\author{Thibaut Devolder}
\email{thibaut.devolder@u-psud.fr}
\affiliation{Universit\'e Paris-Saclay, CNRS, Centre de Nanosciences et de Nanotechnologies, 91120, Palaiseau, France}

\date{\today}                                           
                       
%
%
\begin{abstract}
Many recent papers report on the interest of spin waves for applications. This paper revisits the propagating spin wave spectroscopy when using inductive transceivers connected to a vector network analyzer. The spin wave conduit can be made of a non-reciprocal material. The formalism offers a simple and direct method to understand, design and optimize devices harnessing propagating spin waves, including when a unidirectional energy flow is desired.\\
The concept of the mismatch of helicity between the spin wave and the magnetic field radiated by antennas is first clarified. Owing to the form of the susceptibility tensor reflecting the precession ellipticity, there exists specific orientations of the wavevector for which a perfect helicity mismatch is reached. The spin waves with this orientation and this direction of wavevector are "dark" in the sense that they do not couple with the inductive antenna. This leads to single-sided wavevector generation, that should not to be confused with a unidirectional emission of energy.  \\ 
A method to calculate the antenna-to-antenna transmission parameter is then provided. 
Analytical approximations are then applied on situations that illustrate the respective role of the direction of the spin wave wavevector versus that of the group velocity. The often-encountered cases of spin waves possessing either a  V-shaped or a flat dispersion relation are revisited. These reciprocal dispersion relations lead to amplitude non-reciprocity because of the helicity mismatch phenomenon. Conversely, for spin waves with a line-shaped dispersion relation, an unforeseen quasi-unidirectional emission of spin waves occurs, which is of interest for applications. This situation can be obtained when using the acoustical spin waves of synthetic antiferromagnets when the wavevector is close to parallel to the applied field. We finally show that this configuration can be harnessed to design reconfigurable frequency filters.
\end{abstract}

\maketitle

Spin waves are the resonant low energy disturbances of the ferromagnetic order. First implemented 20 years ago \cite{bailleul_propagating_2003} in frequency domain, the propagating spin wave spectroscopy is now a widely used technique \cite{demokritov_spin_2017} either as a research tool to access fundamental properties like the spin polarization in metals \cite{vlaminck_current-induced_2008} and the spin wave band structure \cite{devolder_measuring_2021}, or for various practical applications \cite{chumak_magnon_2015} including novel computing schemes \cite{csaba_perspectives_2017, talmelli_reconfigurable_2020, korber_pattern_2023} or signal filtering \cite{qin_nanoscale_2021}. Spin waves can be generated or amplified as soon as a torque is applied to the magnetization: spin-orbit torques \cite{gladii_spin_2016}, magneto-elastic torques \cite{thevenard_surface-acoustic-wave-driven_2014}, etc. Since the excitation of spin waves by Zeeman torques and the detection of spin waves by classical induction using simple inductive antennas is the reference method to which other schemes are most often compared \cite{talmelli_spin-wave_2018}, it is of utmost importance to describe exactly this process with a transparent formalism. 

Unfortunately the published descriptions of inductive propagating spectroscopy are incomplete. Some studies focus on the emission \cite{dmitriev_excitation_1988, schneider_phase_2008} of spin waves in specific materials \cite{stancil_theory_1993}. Some others use models customized for a specific configuration \cite{vlaminck_spin-wave_2010, demidov_excitation_2011, vanderveken_lumped_2022}. The studies sometimes disregard one component of the exciting field \cite{ciubotaru_all_2016} and/or of the spin wave stray field \cite{devolder_measuring_2022}. Finally they often make strong assumptions on the material properties \cite{sushruth_electrical_2020} to get physically sound formulas of very relevant practical interest \cite{ciubotaru_all_2016, collet_spin-wave_2017, talmelli_reconfigurable_2020} but of limited validity because their prediction accuracy is difficult to anticipate. Besides, these models cannot be applied to systems in which the spin waves exhibit a strong non-reciprocity, such as the dipolarly-coupled bilayers \cite{qin_nanoscale_2021} or the synthetic antiferromagnets (SAFs) \cite{verba_wide-band_2019, franco_enhancement_2020, gallardo_reconfigurable_2019, ishibashi_switchable_2020, millo_unidirectionality_2023}. 

A noticeable exception is the recent work of Weiss et al. \cite{weiss_excitation_2022} in which a comprehensive self-consistent theory is developed, including almost all the physical phenomena present in a propagating spin wave spectroscopy experiment. While this theory may be regarded as complete, it relies on a heavy mathematical formalism which hinders the physical understanding. Besides, the model must be implemented numerically which prevents from solving the inverse problems. The available hierarchy of models is de facto not adequate to fully answer basic questions such as: how to design a transceiver that generates spin waves with positive wavevectors only ? Or how to design a system that generates spin waves carrying energy in a unique direction ? 

In this work, we substantially upgrade the previous versions of a formalism \cite{sushruth_electrical_2020, devolder_measuring_2021} to provide a comprehensive but still simple and efficient method to model and understand inductive devices harnessing propagating spin waves. We provide physically transparent formulas at the cost of minor approximations. This formalism enables to discern the unique possibilities offered by systems whose spin waves have a linear dispersion relation: the possibility for a unidirectional flow of the spin wave energy, and to additionally tailor its frequency response for frequency filtering.

The paper is organized as follows. Section \ref{TheExcitationProblem} revisits an apparently simple problem: the magnetic field generated by a single-wire antenna and how it excites magnetization precession. A reciprocal space analysis leads to a proper definition of the concept of helicity match/mismatch.
We then analyse the stray field emitted by a spin wave texture and how it couples to a wire antenna used as an inductive detector. We evidence that some spin waves do not emit stray field on one side of the film, such that inductive detection is rigorously blind to these "dark" modes (section \ref{TheDetectionProblem}). We then combine emission, propagation and detection to issue the formalism enabling the calculation of antenna-to-antenna transmission coefficients (section \ref{TheTransmissionProblem}). \\
In the second part of the paper, we apply this formalism to paradigmatic situations. Section \ref{LinDisp} shows the results for spin waves possessing a line-shaped dispersion relation and a diagonal susceptibility tensor. Section \ref{lemma} reports a useful lemma that is required to account for other dispersion relations. Section \ref{Vdisp} consider $\vee$-shaped dispersion relations, still with a diagonal susceptibility tensor. The section \ref{FullSusceptTensor} accounts for the full susceptibility tensor; section \ref{AntiDiag} identifies to what extent this modifies the antenna-to-antenna transmission for all the previously considered dispersion relations. Finally, sections \ref{discussion} and \ref{Applications} discuss the results and propose how to combine the single-sided wavevector emission together with the unidirectional energy flow, in order to design novel frequency filters. The appendix gathers calculation details as well as the specific case of a flat-shaped dispersion relation.

\section{The excitation problem} \label{TheExcitationProblem}
Let us first revisit the apparently simple problem of how the magnetic field generated by a single-wire antenna excites the magnetization of a film underneath.
\subsection{Antenna field in real space}
\begin{figure}
\begin{center}
\hspace*{-0.2cm}\includegraphics[width=8 cm]{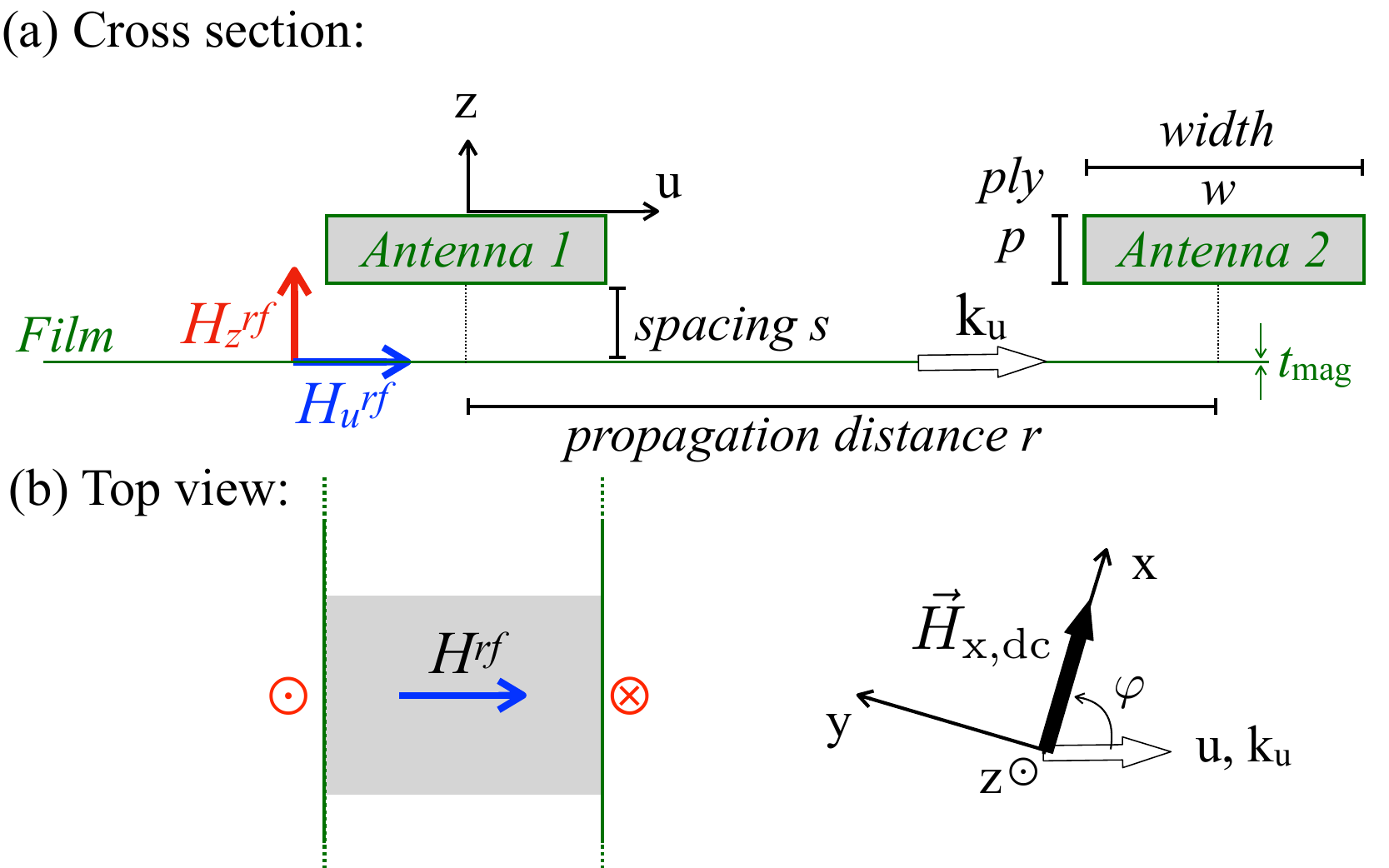}
\caption{(a) Cross section view of the geometry assumed for a propagating spin wave spectroscopy experiment, with the definitions of the main geometrical parameters. (b): Top view with the orientation of the static field $\vec H_{x, dc}$ and the spin wave wavevector $\vec{k}$ with mutual orientation $\varphi=\{\vec{k}, \vec H_{x, dc} \}$ .}
\label{AntennaGeometry}
\end{center}
\end{figure}

We consider a single-wire antenna whose geometry is described in Fig.~\ref{AntennaGeometry}(a). It features a rectangular cross section with a width $w$ along the $\vec u$ axis and a thickness $p$ 
along the $\vec z$ axis, placed at a spacing $s$ above the mid plane of the magnetic film. In the following, we use the $\{u, v, z\}$ coordinates when referring to the antennas, which will couple to spin waves of wavevectors $k$ oriented in the direction $u$. The system is assumed invariant (and infinite) in the $v$ direction. We will use the $\{x, y, z\}$ coordinate when dealing with magnetization dynamics, with $x$ being the direction of the static applied field $\vec H_{x, dc}$. We define the angle $\varphi$ between the spin wave wavevector $\vec k$ and the applied field $\vec{H}_{x, dc}$ [see Fig.~\ref{AntennaGeometry}(b)]. The antenna carries an rf current $I^{rf}$. 

Before calculating the spatial profile of the magnetic field generated by the antenna, it is worth exploiting the symmetries of our geometry. Indeed in all situations  in which (i) the magneto-static problem is invariant in the third direction (here invariant along $\vec v= \vec {e}_z \times \vec u$) and (ii) the sources of the magnetic field, i.e. the electrical currents and the magnetic moments are all in the half plane below or above (in the sense of the $z$ coordinate) the considered position $(u, z)$ of interest,  the two components of the antenna field are mutually linked \cite{mayergoyz_2-d_2006} by a Hilbert transformation with respect to the variable $u$:
\begin{equation}
\left\{
\begin{array}{l}
H_z^{rf}(u, z) = - \mathcal {HT} \{H_u^{rf}(u)\}   \textrm{sgn}(z) \\
H_u^{rf}(u, z) =  ~\,\mathcal {HT} \{H_z^{rf}(u)\} \,\textrm{sgn}(z), \\
\end{array}
\right.
\label{components}
\end{equation}
where "sgn" is the signum function and $\mathcal {HT} \{f(\Omega)\}$ denotes the Hilbert transform of the function $f$ of variable $\Omega$. In our case, the sources of the magnetic field (i.e. the antenna) are above the positions of interest where the field is evaluated. One thus takes sgn(z)$<0$ when considering a position $\{u, z\}$ within the magnetic film.

A classical integration of Biot and Savart law provides the in-plane field $H_u^{rf}(u, z)$ and the out-of-plane field $H_z^{rf}(u, z)$ that the antenna generates\footnote{We neglect the eddy currents that the antenna field generates within the magnetic film. Should the film be conductive, eddy current would partially screen the $H_z^{rf}(u, z)$ field and would smoothen its spatial profile. The sources of the magnetic field would not be located solely above the magnetic film, so that Eq.\ref{components} would no longer strictly hold. This alteration of $H_z^{rf}(u, z)$ would quantitatively (but not qualitatively) change Eq.~\ref{HmM}}. Examples of the field profiles are reported in Fig.~\ref{AntennaFieldsPlot}(a). The expressions of the two field components are heavy and therefore placed in the appendix (see Eq.~\ref{HxdirectSpace1} to \ref{HzdirectSpace2}). For a didactic purpose, we first use the limit of these fields in the case of an ultrathin antenna $(p \rightarrow 0_+)$ carrying the same total current. 
In that limit the two components of the antenna field simplify to\footnote{
Eq.~\ref{hrfx} was checked to be compatible with Maxwell-Ampere law: its circulation about a rectilinear contour making a half turn around the antenna is $\int_{-\infty}^{+\infty} H_u^{rf}(u,z,~p\rightarrow 0) du = \frac{1}{2} I ~\textrm{sgn}(z)$. Besides, Eq.~\ref{hrfz} converges asymptotically to the well-know law: $\forall z$, $h_z^{rf}(u \rightarrow \infty,~p\rightarrow 0) = I / (2 \pi u)$.}
:
\begin{equation} 
\begin{split}
& \mu_0 H_u^{rf}(u,z,~p\rightarrow 0)=  \\ 
&\Sigma \left[\tan ^{-1}\left(\frac{w-2 u}{2 z}\right)+\tan ^{-1}\left(\frac{w+2 u}{2 z}\right) \right] 
\label{hrfx}
\end{split}
\end{equation}
and
\begin{equation}
\mu_0 H_z^{rf}(u,z,~p\rightarrow 0)=\frac{\Sigma}{2}\ln
   \left(\frac{\left(u-\frac{w}{2}\right)^2+z^2}{\left(u+\frac{w}{2}\right)^2+z^2}\right) \label{hrfz} 
\end{equation}
where $\Sigma=\frac{1}{2 \pi} \frac{\mu_0 I^{rf}} {w}$ is scaling factor that has the dimension of a flux density, in Tesla. Eq. \ref{hrfx} and Eq. \ref{hrfz} can be checked \footnote{This can be checked by using Eqs.~\ref{hrfx} and \ref{hrfz} and noting that 
$\mathcal {HT} \{\tan^{-1}(\Omega)\}= \frac{1}{2} \ln(1+\Omega^2)$ and using the time-reversal property of the Hilbert transform (i.e.  $\mathcal {HT} \{f(-\Omega)\}= - \mathcal {HT} \{f\} (-\Omega) $).} to comply with Eq.~\ref{components}. 
The Hilbert transformations in Eq.~\ref{components} means that the two components of the antenna field have correlated spatial variations. Their spectrum in the reciprocal space direction $k$ are thus also intimately related.

\begin{figure}
\begin{center}
\hspace*{-0.2cm}\includegraphics[width=8 cm]{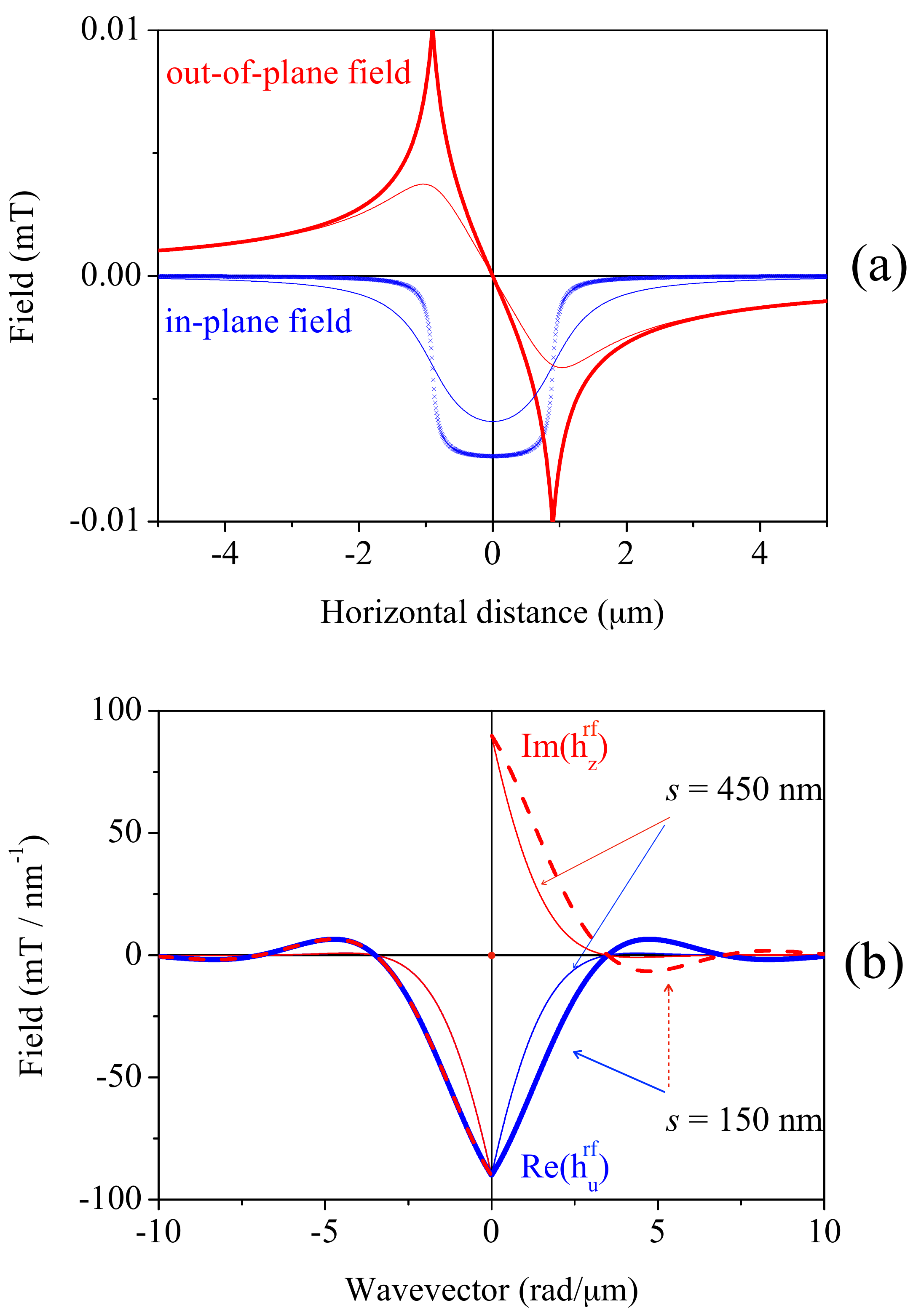}
\caption{(a) Real space and (b) reciprocal space profile of the antenna field along a film placed below the antenna ($z<0$). Antenna thickness and width: $h=0.16~\mu\textrm{m}$ and $w=1.8~\mu\textrm{m}$, antenna above film with two different spacings $s=0.15-0.45~\mu\textrm{m}$, rf current $I^{rf}=1$ mA, or equivalently -13 dBm dc dissipated in an antenna of hypothetical impedance 50 $\Omega$.}
\label{AntennaFieldsPlot}
\end{center}
\end{figure} 

\subsection{Antenna fields in reciprocal space} 
We use the unitary Fourier transform convention adapted for propagating waves: for any function $f$, we define its spatial inverse Fourier transform as $\tilde{f}(k, t) = \frac{1}{\sqrt{2 \pi}}\int_{-\infty}^{\infty} F(u, t) e^{i[\omega t - k u]} du $. We use the tilde overscript "$\tilde{~~}$" to recall that a quantity has a complex value. Using the properties of the Fourier transforms of two functions linked by a Hilbert transformation (Eq.~\ref{components}), we find that the inverse Fourier transforms $\tilde{h}_u^{rf}(k) = \mathcal{F}\{ H_u^{rf}(u) \}$ and $\tilde{h}_z^{rf}(k) = \mathcal{F}\{ H_z^{rf}(u) \}$ of the two components of the antenna field obey for any antenna thickness: 
\begin{equation}
\left\{
\begin{array}{l}
  \tilde{h}_z^{rf}(k) = ~\, i ~\textrm{sgn}(k z)~ h_u^{rf}(k) \\ 
h_u^{rf}(k) = - i ~\textrm{sgn}(k z)~ \tilde{h}_z^{rf}(k)  \\
\end{array}
\right.
\label{spectra} 
\end{equation}
Note that Eq.~\ref{spectra} was also derived in refs.~\cite{dmitriev_excitation_1988, schneider_phase_2008} and the importance of the $k$-odd character of $\tilde{h}_z^{rf}(k)$ was highlighted.
 
\begin{figure}
\begin{center}
\hspace*{-0.2cm}\includegraphics[width=8 cm]{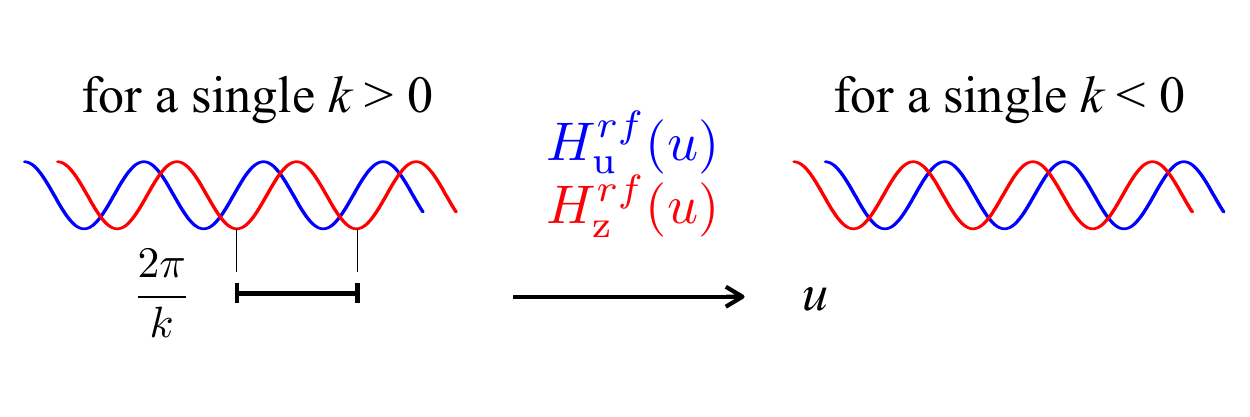}
\caption{Spatial profiles of the two components of the rf field of an antenna when filtered at a single wavevector. }
\label{SingleWaveVectorFieldProfile}
\end{center}
\end{figure}

We emphasize that Eq.~\ref{spectra} means that the two components of the antenna field have the \textit{same} power. Neglecting one component of the antenna field would be an error.
Eq.~\ref{spectra} also means that for each wavevector $k$, the corresponding antenna field is circularly polarized with respect to the $u$ direction, with an helicity sign set by the sign of the product $k z$.  For each $k$, the spatial pattern of the k-monochromatic $\tilde{h}_u^{rf}$ is exactly the same as of the k-monochromatic $\tilde{h}_z^{rf}$ but displaced by a distance of $\frac{\pi} {2 |k|} \textrm{sgn}(k z)$ in the $u$ direction, as sketched in Fig.~\ref{SingleWaveVectorFieldProfile}.

By performing an inverse Fourier transformation 
on Eq.~\ref{hrfx}, we can find the (real-valued) spectrum of the antenna field in the limit of $p \rightarrow 0$:
 \begin{equation}
{h}_u^{rf}(k, p \rightarrow 0) = \frac{I}{2 \sqrt{2 \pi }} { ~\textrm{sinc} \left(\frac{k w}{2}\right)} e^{-\left| k z \right| }\textrm{sgn}(z)  \label{swsh0}
\end{equation}
The "sinc" in the previous expression recalls that the ${h}_u^{rf}(u)$ in real space ressembles a rectangular window and matches with it when $p \rightarrow 0$ and $z \rightarrow 0^+$.

For an antenna of arbitrary thickness $p>0$, a simple integration provides:  
 \begin{equation}
 \begin{split}
&h_u^{rf}(k, p>0)= \\ 
&\frac{I}{2 \sqrt{2 \pi }}          \textrm{sinc} \left(\frac{k w}{2}\right) 
\left(\frac{1-e^{-h \left| k\right|}    }    {    h \left| k\right| } \right) e^{-\left| k s\right| }\text{sgn}(z)  \\
\label{AntennaField}
\end{split}
\end{equation}
where the $\textrm{sgn}(z)$ should be taken as $-1$ since the antenna is above the film.
Examples of the field profiles in reciprocal space are reported in Fig.~\ref{AntennaFieldsPlot}(b) for representative spacings\footnote{Note that the $k=0$ case is pathological in Eq.~\ref{spectra} and their consequences from Eq.~\ref{swsh0} and \ref{AntennaField}: we have $\tilde{h}_z^{rf}(k=0)=0$ as this value should be the $u$-integral of the ${h}_z^{rf}(u)$ field that has $u$-odd symmetry [see Fig.~\ref{AntennaFieldsPlot}(a)]. Mathematically, this singularity comes from the fact that the function ${h}_z^{rf}(u)$ is not absolutely integrable as it asymptotically decays like $1/u$ as $u \rightarrow \pm \infty$. Physically, this singularity arises because the current in the device [Fig.~\ref{AntennaFieldsPlot}(a), inset] flows from its source at $v=-\infty$ to $v=\infty$ without ever circulating back to its source; this singularity would be smoothed out in any real configuration because the current must return back by some path at a finite distance $u_0$, leading to a $1/u^2$ decay of $H_z$ for $|u| \gg u_0, h, w$.\label{PathologicalNote}}. 

\subsection{Linear, circular or elliptical polarization of the antenna field in the coordinate system of the magnetization} \label{HmMsection} 
For each wavevector, Eq.~\ref{spectra} means that the antenna field is circularly polarized when looked at in the coordinate system  $\{\vec u, \vec v, \vec z \}$ common to the antenna and to the spinwave wavevector. It is generally not the case in the frame of the film's magnetization, since this frame is rotated by an angle  $\varphi$ (see Fig.~\ref{AntennaGeometry}). 

When $\varphi=0$ deg., we have $\{\vec x, \vec y, \vec z \} = \{\vec u, \vec v, \vec z \}$. In many situations magnets respond only to the rf fields that are in the in the $yz$ plane transverse to the applied field, where the sole non-zero component is $h_z^{rf}$. Such magnets are thus excited by an rf field that seems \textit{linearly polarized} along $z$. 

In contrast when $\varphi=90$ deg., we have $\{\vec x, \vec{y}, \vec z \} = \{\vec v, -\vec u, \vec z \}$ and the antenna field circularly polarized in the $uz$ plane can be described with an opposite \textit{circular polarization} in the $yz$ plane. 
Using Eq.~\ref{spectra}, we have that intermediate angles $\varphi$ provide \textit{elliptically polarized} rf fields of components: 
\begin{equation}
\left\{
\begin{array}{l}
\tilde{h}_z^{rf}(k)= \,i\,~ h_u^{rf}(k)\, \textrm{sgn}(k z) \\
h_x^{rf}(k)= ~~\, h_u^{rf}(k) \cos(\varphi)   \\
h_y^{rf}(k)= - h_u^{rf}(k) \sin(\varphi) \\
\end{array}
\right.
\label{rotatedField}
\end{equation}

\subsection{Helicity mismatch to suppress the magnetization response} \label{HmMsection} 
The different components of the rf field (Eq.~\ref{rotatedField}) can either work together, or against each other to generate spin waves, depending on whether the helicities of the rf field and of the magnetization precession match or not. Since the helicity of the antenna field depends on the sign of $k$ (see Eq.~\ref{spectra}), the degree of matching depends on the spin wave propagation direction. This is generally seen as a difference of the efficiency of the antenna to excite the $\pm k$ counterpropagating spin waves \cite{schneider_phase_2008}, leading to some amplitude non-reciprocity.

To achieve a physically-transparent description of this "helicity mismatch" effect, let us look at the high frequency susceptibility tensor $\bar{\bar\chi}(k)$ of the total moment of a magnet system. For a didactic purpose, we temporarily disregard the fact that $\bar{\bar\chi}(k)$ and $\bar{\bar\chi}=\bar{\bar\chi}(k=0)$ slightly differ\footnote{They have generally different resonance frequencies and linewidths but their shapes have a high degree of similarity}. With an applied field along $x$, $\bar{\bar\chi}$ is generally finite only the $\{y, z\}$ plane. In this plane it reads:
\begin{equation}
\bar{\bar\chi} = \chi_\textrm{max} \left(
\begin{array}{cc} 
 1  & -i \epsilon   \\
 i \epsilon    & \epsilon ^2 \\
\end{array}
\right)
\label{chi}
\end{equation}
where $\chi_\textrm{max} \in i \mathbb{R}^-$ is the value of the $\chi_{yy}$  term of the susceptibility at resonance, and $\epsilon$ is the precession ellipticity.

The magnetization response $\tilde{m}^{rf}(k) = \bar{\bar {\chi}} \, \tilde{h}^{rf}(k) $ follows from Eq.~\ref{rotatedField} and \ref{chi}. The response vanishes in the "perfect" helicity mismatch (HmM) case. This happens for specific field directions $\varphi$, which read simply: 
\begin{equation}
\sin(\varphi_{HmM})= \epsilon ~\textrm{sgn}(k z) 
\label{HmM}
\end{equation}
For these angles, the antenna field only excites one of the two directions of wavevector\footnote{Note that an expression similar to Eq.~\ref{HmM} was derived in ref.~[\onlinecite{fripp_spin-wave_2021}] for a different transducer [\onlinecite{au_resonant_2012}] and $\varphi=\pi/2$.}. These HmM angles depend on the precession ellipticity $\epsilon(k, H_{x,dc})$ only, which is weakly field and wavevector dependent. This single-sided wavevector generation is not to be confused with the generation of a unidirectional energy flow. This point can be very important when looking at angular variations in propagating spin wave spectroscopy. We will come back to this point later.

This form of $\bar{\bar{\chi}}$ in which (Eq.~\ref{chi}) the non-diagonal components are shifted in phase by $\pm i$ (i.e. $\pm \pi/2$ radians) with respect to the diagonal elements occurs in many situations \cite{stancil_theory_1993}. Noteworthy, this includes the total moment of a synthetic antiferromagnet (SAF) at its acoustical resonance \cite{millo_unidirectionality_2023}, which is the material system that we shall harness for the application proposed in section \ref{discussion}. In this case the precession ellipticity is $\epsilon=\frac{\sqrt{H_j}}{\sqrt{H_j+M_s}}$ at low fields, where we have defined the interlayer exchange field $\mu_0 H_{j}=-\frac{2 J}{ {M_{s}} {(t_\textrm{mag}/2)}}$ with $J$ being the interlayer exchange coupling and $t_\textrm{mag}$ the total thickness of the SAF. In numerical applications, we shall take the material parameters of a CoFeB/Ru/CoFeB SAF from ref.~[\onlinecite{mouhoub_exchange_2023}], i.e  $t_\textrm{mag}=34$ nm, $J$=-1.7 mJ/m$^2$, $\mu_0 M_s=1.7$ T. This leads to a precession ellipticity $\epsilon \approx 0.3$ for the acoustical spin wave branch, so that the perfect helicity mismatch angles are $\varphi_{HmM}$=16 or 164 degrees for $k<0$ and equivalently 163 and 343 degrees for $k>0$.

\section{The detection problem} \label{TheDetectionProblem}
\subsection{From the dynamic magnetization of the spin wave to its stray fields} 
\subsubsection{Formalism} 
We now consider a spin wave of wavevector $k$ with dynamical components $m_{u, v, z}(u, t)= \textrm{Re} \left(\tilde{m}_{u, v, z} e^{i[\omega t - k u]} \right) $ with complex amplitudes $\{\tilde m_u, \tilde m_v=0, \tilde m_z\}$ (in units of A/m). From now on we postulate that the magnetization can be replaced by its thickness-averaged value. The stray field will be simpler to express in the frame of the magnetic film, so that we now take the $z'=0$ plane as the center of the film of thickness $t_\textrm{mag}$. The stray field created by this spin wave can be calculated as in ref.~[\onlinecite{mallinson_one-sided_1973}]. The $\tilde m_u$ generates a volume density $\tilde \rho(u, z')= i k \tilde m_u e^{-i k u}$ of magnetic pseudo-charges within the magnetic body, from which the magnetostatic potential can be deduced by integration. Taking its gradient we get that the in-plane dynamic component $\tilde{m}_u$ of the magnetization generates a stray field outside of the film whose in-plane component is:
\begin{equation} H_u^\textrm{stray, mu}(u, z')= - \tilde m_u e^{-\left|k z' \right|} e^{-i k u} \textrm{sh}\left(\frac{k t_\textrm{mag}}{2}\right) \textrm{sgn}(k),\end{equation}
In parallel, the $\tilde m_z$ component of the spin wave generates a surface density $\tilde \sigma(u, z')=\pm \delta_{z' \mp t/2} \tilde m_z e^{-i k u} $ of magnetic pseudo-charges at the top (+) and bottom surfaces (-) of the magnetic body, from which the corresponding magnetostatic potential can be deduced. This gives another source of in-plane oriented stray field: 
\begin{equation} H_u^\textrm{stray, mz}(u,z')= i~ \tilde m_z e^{-\left|k z' \right|} e^{-i k u} \textrm{sh}\left(\frac{k t _\textrm{mag}}{2}\right) \textrm{sgn}( z') \end{equation}

\begin{figure*}
\begin{center} 
\includegraphics[width=17 cm]{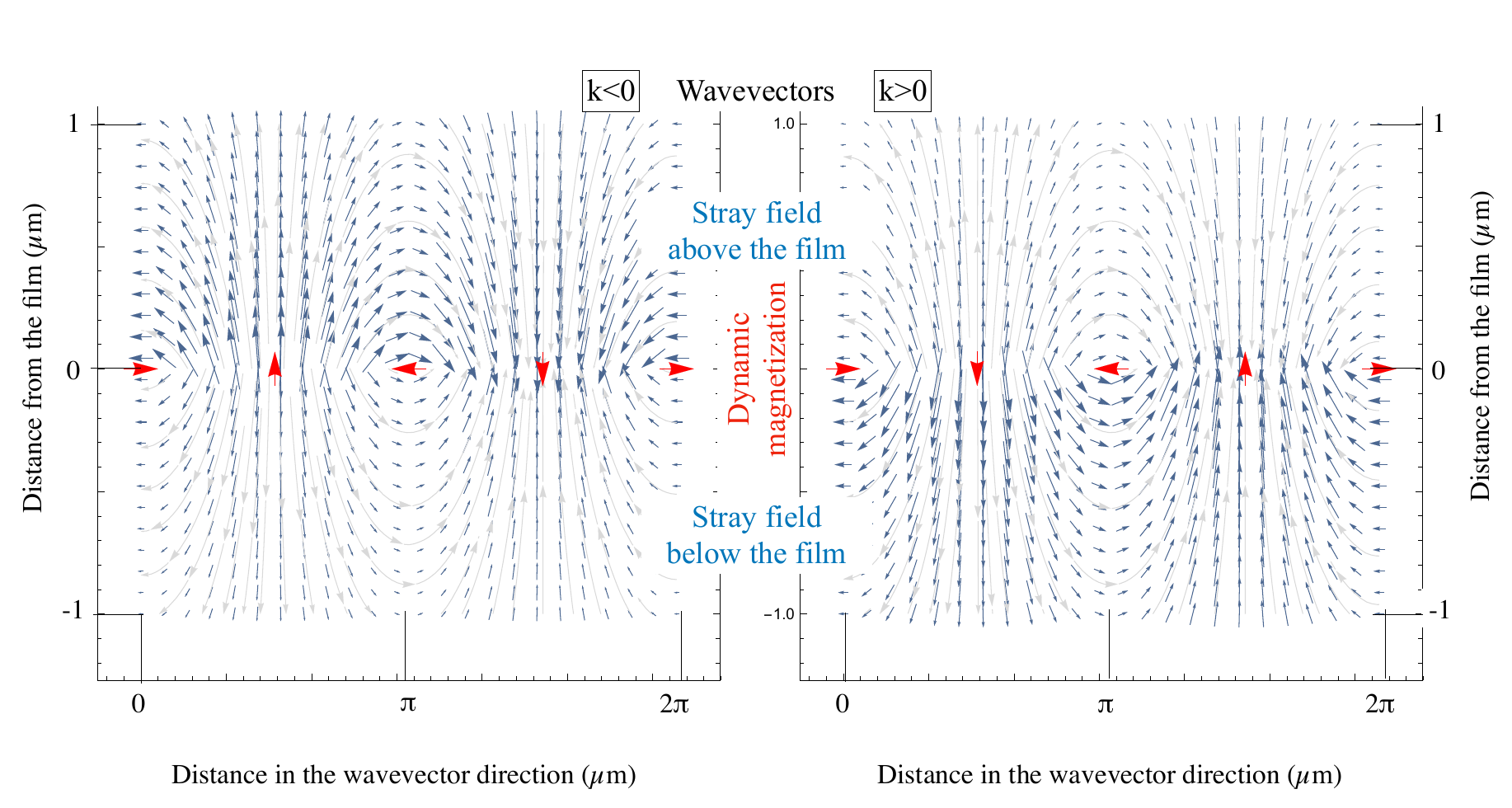}
\centering
\caption{Stray field outside of a thin film for counter-propagating spin waves with ellipticity such that $\tilde m_z= - 0.3\,i\, \tilde{m_u}$, following Eqs.~\ref{Hudemag} and \ref{Hzdemag}. The spin waves have a wavelength of $2\pi /k$ for $|k|=1~\textrm{rad}/\mu$m. Left: for $k<0$. Right: for $k>0$. The red arrows sketch the normalized dynamic magnetic magnetization $\{Re  (\tilde m_u e^{[i (\omega t - k u )}), Re  (\tilde m_z e^{[i (\omega t - k u )}) \}$ inside the film at time 0 for a film of hypothetical vanishing small thickness. The sizes of the blue arrows scale with the amplitude of the stray field (arb. units). }
\label{DemagFieldPattern}
\end{center}
\end{figure*}

For one spin wave of wavevector $k$, the total stray field is the sum of these two contributions:
\begin{equation}
\begin{split}
&H_u^\textrm{stray}(u,z') =  \\ &\Big[ i ~\tilde m_z ~\textrm{sgn}(z')  - \tilde m_u \textrm{sgn}(k) \Big] e^{-\left|k z' \right|} e^{-i k u} \textrm{sh}\left(\frac{k t_\textrm{mag}}{2}\right) 
\label{Hudemag}
\end{split}
\end{equation}
We will see later that this $u$-oriented component of the dynamic stray field is the sole that matters for the inductive detection by the receiving antenna; it is however instructive to also have a look at the other field component. The sources $\tilde \rho$ and $\tilde \sigma$ of the stray magnetic field are now all \textit{below} the position of interest for detection (the antenna), such for the solitary spin wave of wavevector $k$, the spatial profile of the stray field follows now a modified version of Eq.~\ref{spectra}:
\begin{equation}
\tilde {H}_z^\textrm{stray}(u, z') = -\,i\,\textrm{sgn}(k z') \, \tilde{H}_u^\textrm{stray}(u, z')  
\label{Hzdemag}
\end{equation}
which\footnote{This expression could also have been derived directly using $\vec\nabla . \vec{H} = 0$ outside of the film and the fact that the sole space dependence of the stray field components is its periodicity $e^{-i k u}$ and its decay $e^{-\left|k z' \right|}$. } holds out of the film (i.e. for $|z'|\geq \frac{t_\textrm {mag}}{2}$) and for a single spin wave of wavevector $k$. An example of the stray field profile associated to a a single spin wave is given in Fig.~\ref{DemagFieldPattern}. We emphasize that owing to the minus sign in Eq.~\ref{Hudemag}, the amplitude of the stray field on one side of the film is stronger than on the other side. This is particularly important when intending to detect the spin wave inductively. 

\subsubsection{Single-sided stray field cases versus helicity mismatch} 
It is interesting to scrutinize when Eq.~\ref{Hudemag} and \ref{Hzdemag} vanish for one sign of $kz'$, because this means that the spin-wave emits flux in one half $z$-space only, either above or below the film, in a manner similar to the "fridge magnet" configuration \cite{mallinson_one-sided_1973} approximated in Halbach linear arrays of permanent magnets \cite{johns_mobile_2015}. As well known, a single-sided stray field happens every time the two components of the dynamic magnetization are related by a Hilbert transformation \cite{mallinson_one-sided_1973}, or equivalently when Eq.~\ref{Hzdemag} applies for all the wavevectors present in the magnetization texture.

Going back to the frame $\{u, z\}$ that is the one relevant for the inductive detection, the components of any spin wave $\{ \tilde m_x, \tilde m_y, \tilde m_z\}$ can be rewritten as $\tilde m_u=- \tilde m_y \sin \varphi$ and $\tilde m_z$. Eq.~\ref{Hudemag} and \ref{Hzdemag} can be used to get the dynamical components of the stray field in the projected in the plane relevant for inductive detection. These components vanish for the no-inductive-detection (nid) condition, defined as:
\begin{equation}
\sin{\varphi_\textrm{nid} }=- i \frac{\tilde m_z}{\tilde m_y} \textrm{sgn}(k z')
\label{nid}
\end{equation}
For an antenna placed above the magnetic film ($z'>0$ and $z<0$), the conditions of perfect helicity mismatch (Eq.~\ref{HmM}) for which  a spin-wave of wavevector $k$ is strictly not responding to the antenna field, and the condition (Eq.~\ref{nid})  for which the same spin wave cannot be detected by the same antenna, are identical.

This intuitive conclusion can be rephrased in a simple rule: \textit{if an inductive antenna cannot excite a spin-wave, it also cannot detect inductively this spin wave.} This rule is not specific to inductive antennas, and was for instance also observed for transducers based on nanomagnet resonators  \cite{au_resonant_2012, fripp_spin-wave_2021, wang_nonreciprocal_2021}.

\subsection{From the stray field of the spin wave to the electromotive force} 
We now need to get the electromotive force (e.m.f.) that this stray field induces on the receiving inductive antenna. 
For a start, we consider an ultrathin and ultranarrow antenna  (i.e. $p, w \rightarrow 0$) . Using the Maxwell-Faraday equation which reads $\vec{\nabla}\times \vec{E}= - \frac{\partial \vec B}{\partial t}$, the time variation of the spin wave stray field generates an electric field whose component in the $\vec v$-direction is: 
 \begin{equation}
 \vec{E}_v (u, z', p=w=0) = - i \omega \int_s^{\infty}\tilde{H}_u^\textrm{stray}(u, z') dz' \end{equation}
Considering a finite size antenna requires to average this expression over $z'$ spanning from $s$ to $s+p$ and over $u$ spanning from $-\frac w 2$ to $\frac w 2$. This gives a mutual inductance that scales with the e.m.f. per unit length on the receiving antenna:  
 \begin{equation}
 \begin{split}
&\frac{1}{i \omega} \frac{\partial \textrm{e.m.f.}}{\partial v} =  \textrm{sinc}(k w /2) \times  \Big[  \tilde m_u - i ~\tilde m_z ~\textrm{sgn}(k z')  \Big]  \\& \times
  \left(\frac{1-e^{-p \left| k\right|}    }    {    p \left| k\right| } \right) e^{-\left| k s\right|}
.  \frac{\textrm{sh}\left(\frac{k t_\textrm{mag}}{2}\right)}{k}
\label{fem}
\end{split}
\end{equation}
In practice a typical experiment is performed with wavevectors bounded by the antenna width, i.e. $|k| < \frac {2 \pi} w$, on a device whose geometry complies with $w \gg  s, p, t_\textrm{mag}$, such that the second line of Eq.~\ref{fem} can be reasonably approximated by $\frac{t_\textrm{mag}} {2}$.
 
\section{The transmission problem} \label{TheTransmissionProblem} 
\subsection{Contribution of the spin waves to the antenna-to-antenna scattering matrix} 
Let us calculate the antenna-to-antenna transmission parameter as would be measured by a vector network analyser (VNA) at an applied frequency $\frac \omega {2 \pi}$. This frequency is a priori distinct from the uniform resonance frequency $\omega_0$. 
We start by evaluating the response at a single wavevector $k$. We express the antenna field at the emitter (Eq.~\ref{AntennaField} and \ref{spectra}) in the frame of the applied dc field, i.e. the $\{x, y, z\}$ frame. For an antenna placed above the film, this stimulus is $\bar{h} h_u^{rf}(k,~z)$ with $h_u^{rf}$ is from Eq.~\ref{AntennaField} and: 
\begin{equation} \bar{h}=\big(\cos(\varphi), -\sin(\varphi), - i~ \textrm{sign}(k)\big) 
\label{h} \end{equation} 
We then evaluate the magnetization response below the emitter antenna using $\tilde{m}(k) \big{\vert}_{u=0} = \bar{\bar {\chi}} \,. (\bar{h} {h_u}^{rf}(k))$. 
Note that the dynamic susceptibility $\bar{\bar{\chi}}_{(\omega,k)}$ has to be expressed for the wavevector and frequency currently under investigation, and in the $\{x, y, z\}$ frame of the magnetic field.

This dynamic response is propagated towards the receiving antenna at the position $u=r$ by a multiplication by the phase accumulation scalar term:
$$ \tilde{m}(k) \big{\vert}_{u=r} = \tilde{m}(k) \big{\vert}_{u=0}  e^{-i kr} $$
Then this dynamic magnetization below the receiving antenna is reprojected in the $\{u, v, z\}$ frame of the antenna to get $\big(\tilde m_u(r), m_v(r), \tilde m_z(r)\big)$, thus enabling the calculation of the e.m.f of the receiving antenna. This expression is simplified once noticing in Eq.~\ref{fem} that the e.m.f. scales like the scalar product between the magnetization and $ \bar h^{*}$, where the star symbol denotes the complex conjugate. The current-in to e.m.f-out transmission parameter as enabled by the spin waves between antenna 1 and antenna 2 reads: 
\begin{equation}
\begin{split}
&\tilde{S}_{21f}(\omega, r) \propto \\ 
&\int_{-\infty}^{+\infty} \Big[ \bar h^{*}.~[ \bar{\bar{\chi}}_{(\omega,k)} . \bar h] \Big]~ e^{-i kr}~ 
\Big( h_u^{rf}(k,z)\Big)^2 dk
\label{S21}
\end{split}
\end{equation}
where we have omitted the numerical prefactor ${2 \sqrt{2 \pi}} \left(\frac{1}{k}\textrm{sh}(\frac{k t_\textrm{mag}}{2})\right)$ that is essentially independent from $k$ but is proportional to $w t_\textrm{mag}$ i.e. the volume of the spin wave conduit. The subscript label "f" refers to the integration over the \textit{full} spectrum (i.e. $k \in ]-\infty, +\infty[$), in opposition to the partial responses to be calculated later in this paper. \\The reflection parameter $S_{11f}$ and reverse transmission parameter $S_{12f}$ can be calculated with the same expression but with $r=0$ and $r<0$. Note also that we write abusively $S_{21}$ and $S_{11}$ quantities that, strictly speaking, would not the ones measured with a VNA, since the VNA scattering parameters would be also functions of additional impedances within the circuit.

\subsection{Transmission in the absence of Gilbert damping}
The most trivial situation is the limit of vanishing Gilbert damping $(\alpha \rightarrow 0)$ but arbitrary dispersion relation. In that case for any applied frequency $\omega$, there is a small number of resonant spin waves of wavevectors $k_0$'s, each with a vanishing linewidth and a diverging susceptibility. \\
When $\alpha \rightarrow 0$ each resonant spin wave contributes to the susceptibility tensor by a term essentially of the form:
\begin{equation}
\left( - i \frac{2\pi}{t_\textrm{mag}}\delta_{k_0}  \pm \frac{1}{t_\textrm{mag}} \frac{1}{k-k_0} \right)\left(
\begin{array}{cc} 
 1  & -i \epsilon(k_0)   \\
 i \epsilon(k_0)    & \epsilon(k_0) ^2 \\
\end{array}
\right) \label{PoleSusceptibility}
\end{equation}
where $\delta$ is the Dirac distribution and the $\pm$ sign is the sign of the group velocity at the considered $k_0$. The contribution of each $k_0$ (i.e. each pole) of the susceptibility to the antenna-to-antenna transmission coefficient can be calculated from the combination of Eq.~\ref{S21}, \ref{PoleSusceptibility} and the residue theorem. Overall, we have:
\begin{equation}   
\begin{split}   
&\tilde{S}_{21f}(\omega, r, \alpha \rightarrow 0) \propto  \\
&- i \frac{4\pi}{t_\textrm{mag}}  \sum_{\textrm{resonant~} k_0s} X_{k0}~ e^{-i k_0 r}~ 
\Big( h_u^{rf}(k_0)\Big)^2  
\end{split}
\label{trivialLimit} 
\end{equation}
where $X_{k0}$ are helicity mismatch numerical terms that depend on the sign of each $k_0$, on $\varphi$ (see Eq.~\ref{h}), on the precession ellipticity and on the peak susceptibility value. Note that Eq.~\ref{trivialLimit} is the intuitive formula used in many places \cite{ciubotaru_all_2016, talmelli_reconfigurable_2020, devolder_measuring_2021, sushruth_electrical_2020, wang_nonreciprocal_2021} when a qualitative understanding is sufficient. It expresses that the frequency dependence of the phase of $\tilde{S}_{21f}(\omega, r, \alpha \rightarrow 0)$ can be used to get the dispersion relation provided a single mode contributes to the transmission parameter\cite{devolder_measuring_2021}. 
In practice, the damping is finite and Eq.~\ref{trivialLimit} has to be tested against more realistic calculations.

\subsection{Particular cases where analytical solutions exist}
Although we will often integrate Eq.~\ref{S21} numerically to get exact results, it is interesting to look at situations for which $\tilde{S}_{21f}(\omega, r)$ can be calculated analytically and lead to physically transparent expressions at the cost of minor approximations. \\

\begin{figure}
\begin{center} 
\includegraphics[width=8.5 cm]{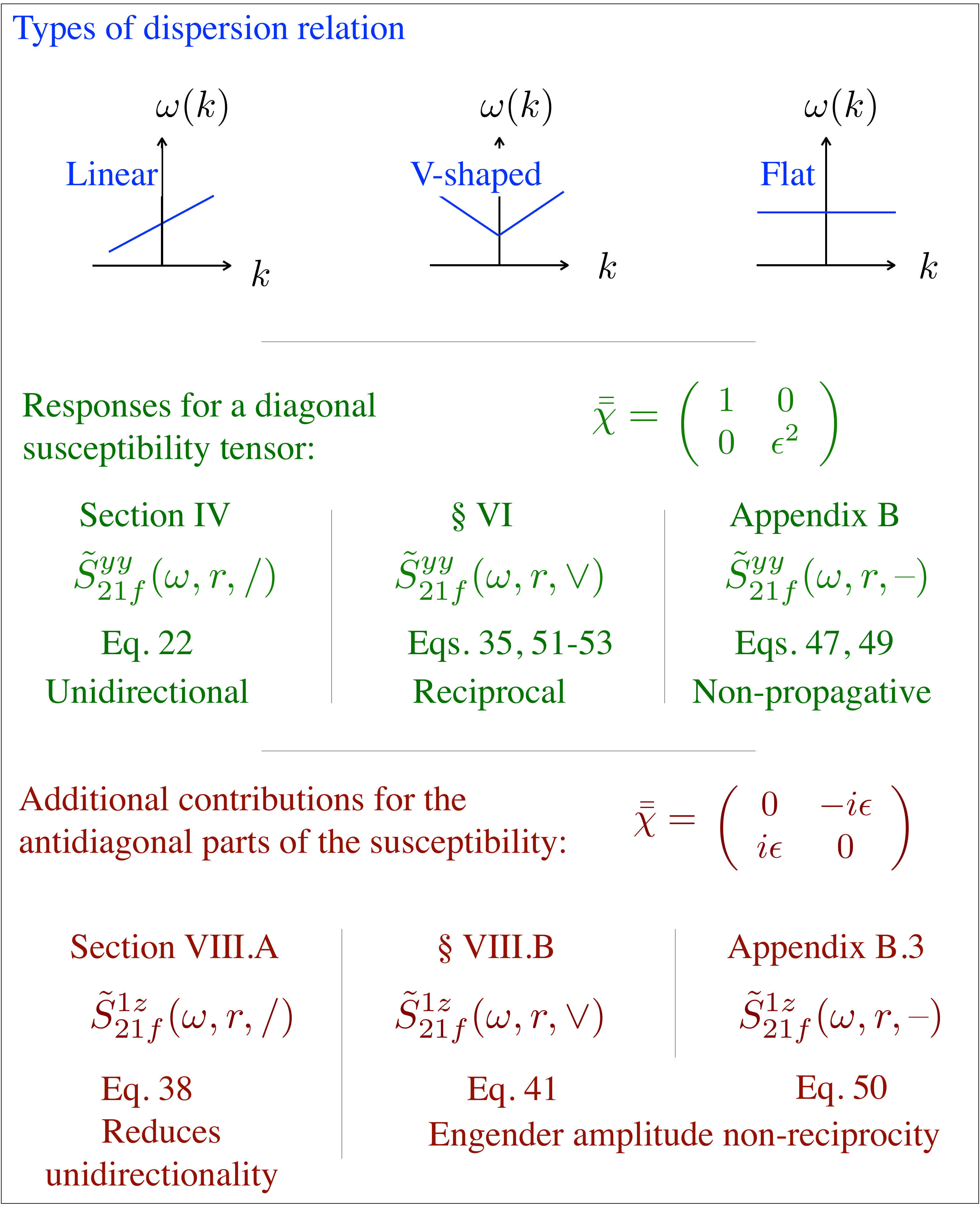}
\centering
\caption{Types of dispersion relations (blue) for which illustrative analytical formulas are derived. Contributions of diagonal elements of the susceptibility: sections, notations, equations and main feature (green).  Idem for the non-diagonal elements of the susceptibility (brown). }
\label{outline}
\end{center} 
\end{figure}

These situations happen when the dispersion relation is line-shaped or broken-linear in the "relevant" wavevector allowed by the antenna geometry (i.e. typically for $k \in [-\frac{2 \pi}{w}, \frac{2 \pi}{w}]$). The analytical calculations require to treat the different terms of $\bar {\bar{\chi}}$ one by one to circumvent the difficulty associated with the sgn($k$) dependence of the $z$ component of $\bar h$ within Eq.~\ref{S21}. We shall sort these contributions in two categories. \\%
We will use the superscript $yy$ in $\tilde{S}_{21f}^{yy}$ for transmission parameters involving a diagonal term of the susceptibility matrix, leading to an integrand of Eq.~\ref{S21} that is independent from sgn($k$). 
Conversely, the susceptibility terms in $\Big[ \bar h^{*}.~[ \bar{\bar{\chi}}_{(\omega, k)} . \bar h] \Big]$ that involve the component $\bar h_z$ exactly once lead to a sgn($k$) dependence of the integrand within Eq.~\ref{S21}. For these terms, we shall use the superscript $1z$ [short for $\bar {\tilde h}_z$ is involved exactly once]. \\
The interesting situations are (Fig.~\ref{outline}):
\begin{itemize}
\item When the dispersion relation is monotonously linear across the relevant $k$ spectrum, and when in addition the magnetization is excited by a diagonal susceptibility term only (section \ref{LinDisp}). \\
The label "/" in  $\tilde{S}_{21f}^{yy}(\omega, r, /)$ will be used to recall the line-shaped nature of the dispersion relation with a positive slope $v_g$, in contrast to the self-understandable other cases displayed in Fig.~\ref{outline}: $\tilde{S}_{21f}^{yy}(\omega, r,  \backslash)$, $\tilde{S}_{21f}^{yy}(\omega, r, \vee)$ and $\tilde{S}_{21f}^{yy}(\omega, r, -)$.
\item When the dispersion relation or the rf field are different for the two signs of $k$ so that two partial responses must be calculated separately (to be referred as $\tilde{S}_{21+}^{yy}$ for $k>0$ and $\tilde{S}_{21-}^{yy}$ for $k<0$, section \ref{lemma}). The lemma of this situation (Eq.~\ref{S21kpos}) will be used extensively in the later calculations.
\item When the dispersion relation is $\vee$-shaped and the magnetization is excited by a diagonal susceptibility term only [to be referred as $\tilde{S}_{21f}^{yy}(\vee)$, section \ref{Vdisp}]. 
\item When the dispersion relation is flat and the magnet is excited by a diagonal susceptibility term only [to be referred as $\tilde{S}_{21f}^{yy}(-)$, appendix \ref{FlatDisp}].
\item Finally, we shall derive the full response, i.e. when all susceptibility terms are included in the response [to be referred as $\tilde{S}_{21f}$, first formally in section \ref{FullSusceptTensor}] then case-by-case in section \ref{AntiDiag} and appendix \ref{FlatDisp}(3)].
\end{itemize}

From now on, we define $\omega_0$ and $\Delta\omega_0$ as the $k$=0 (uniform) resonance frequency and linewidth at the presently applied field. At the presently applied frequency $\omega \neq \omega_0$, there can exist resonant spinwaves. Their wavevectors will be written $k_0$'s. 

\section{Linear dispersion relation and diagonal susceptibility: unidirectional energy flow} \label{LinDisp}
In this section, we assume a perfectly linear dispersion relation $\omega(k) = \omega_0 + v_g k $ with $v_g \neq 0$, valid up to the maximum wavevector compatible with the antenna geometry. In practice, this kind of dispersion relation can arise in dipolarly-coupled bilayer \cite{qin_nanoscale_2021} and in synthetic antiferromagnet \cite{millo_unidirectionality_2023} near $k$=0. In the present section, we assume that the antenna is ultrathin and that it is in direct contact with the film (i.e. $s=p=0$). We also assume that the magnet reacts to the rf field by only one of its diagonal susceptibility terms (e.g. $yy$); therefore we use the notation $\tilde{S}_{21f}^{yy}(\omega, r, /)$.  

At the applied frequency $\omega$, there is a single resonant spinwave, whose wavevector is $k_0=(\omega-\omega_0)/v_g$. The considered susceptibility term can be expanded to first order in $k$ and written in the form\cite{sushruth_electrical_2020}: 
\begin{equation}
\chi(k)=-\frac{i~ {\Delta \omega_0} ~\chi_\text{max}}{2 v_g \left(k-{k_0}+\frac{i}{{L_\textrm{att}}}\right)},
\label{chidek}\end{equation}where $L_\textrm{att}=\frac{2 v_g}{\Delta\omega_0}$ is the attenuation length \cite{gladii_spin_2016}.
Let's first consider a positive group velocity, hence $L_\textrm{att} >0$. 
\subsection{Linear dispersion with a positive group velocity} \label{LinDispYY}
\begin{figure}
\begin{center}
\hspace*{-0.3cm}\includegraphics[width=9.1 cm]{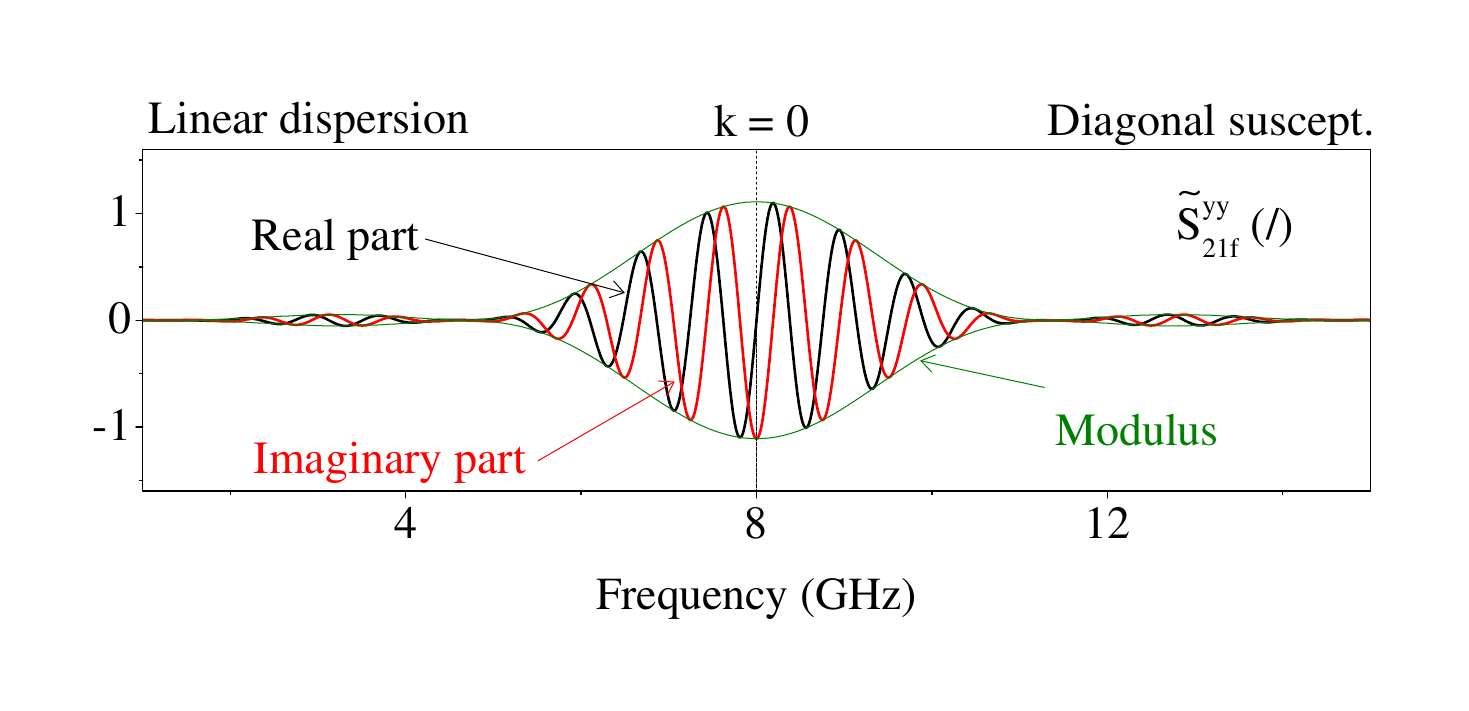}
\caption{Frequency dependence of the antenna-to-antenna transmission parameter $\tilde{S}_{21f}^{yy}(\omega, r=8~\mu\textrm{m}, /)$  for a diagonal susceptibility term and a linear dispersion relation (Eq.~\ref{S21fLinearDisp}). The uniform resonance is $\frac{\omega_0} {2 \pi}$=8 GHz, with $v_g=6$ km/s and $L_\textrm{att}=6~\mu$m. The antennas have zero thickness ($p$=0) and are in direct contact with the film $s=0$. The width of the antennas is $w=1.8~\mu$m.}
\label{AnalyticalS21versusF}
\label{AnalyticalS21versusFlinearDispersion}
\end{center}
\end{figure}

\begin{figure}
\begin{center}
\hspace*{-0.5cm}\includegraphics[width=9.3 cm]{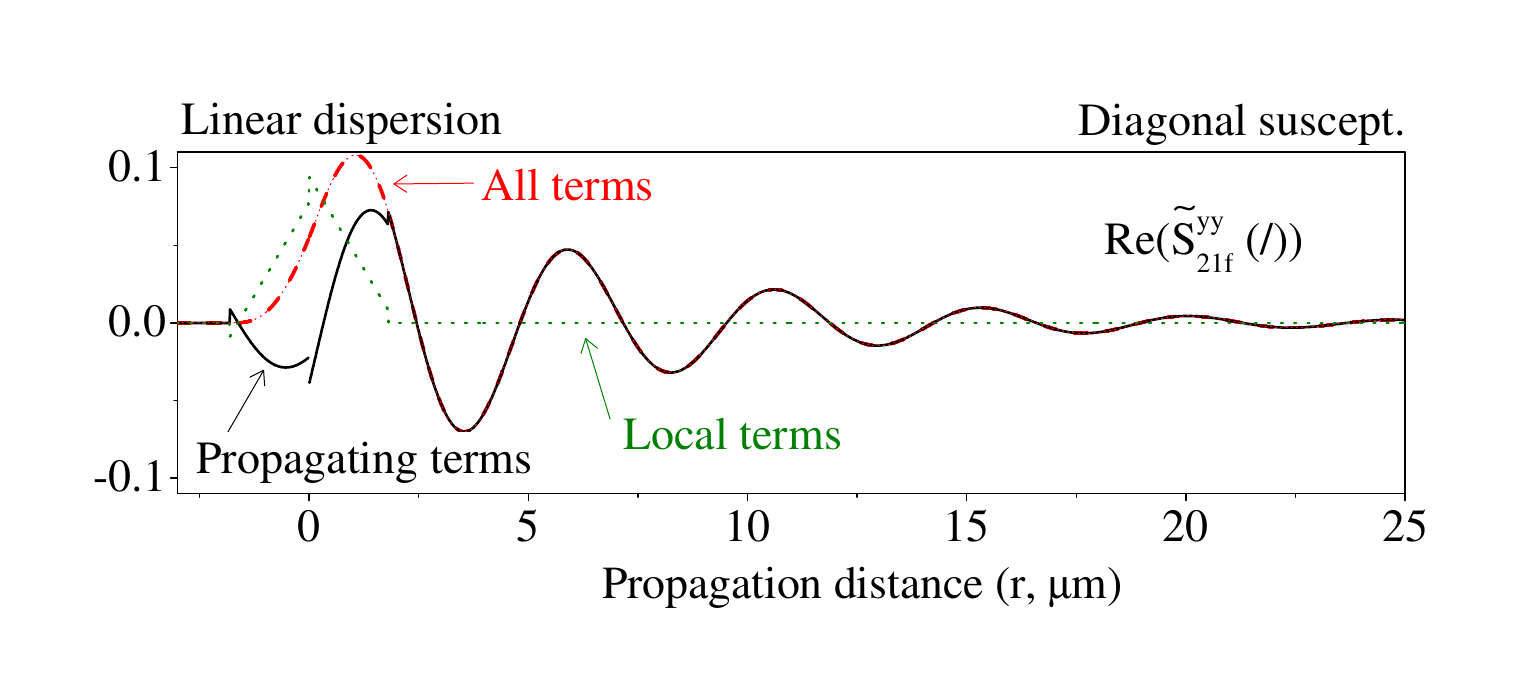}
\caption{Distance dependence of the different contributions to the real part of the antenna-to-antenna transmission parameter $\mathcal{R}e\left(\tilde{S}_{21f}^{yy}(\omega, r, /)\right)$ for a linear dispersion relation and a diagonal susceptibility  (Eq.~\ref{S21fLinearDisp}). The parameters are $\frac{2 \pi v_g}{\omega-\omega_0}=4.7~\mu$m, $L_\textrm{att}=6~\mu$m and $v_g=6$ km/s. 
The antenna geometry is $p=s=0$ and $w=1.8~\mu$m. Note the unidirectional character of the energy flow.}
\label{AnalyticalS21versusR}
\end{center}
\end{figure}

After some algebra and using Eq.~\ref{chidek} one can rewrite Eq.~\ref{S21} in a form containing explicitly propagating ($P$) and non-propagating ($L$, for local) contributions, as well as a common Lorentzian-resonant prefactor $\mathcal{L}(\omega)$:
\begin{equation} \begin{split}
&\tilde{S}_{21f}^{yy}(\omega, r, /) ~~=~~ \overbrace{\sqrt{\frac \pi 2} ~\mathcal{L}(\omega)}^{Lorentzian} ~~\times  \\ &  \Big[\underbrace{P_1(r, \omega)}_{propag.} \underbrace{U(r)}_{unidir.} \underbrace{-2 L(r) + L(r+w) + L(r-w)}_{local}  \Big]
\end{split}
 \label{S21fLinearDisp} \end{equation}
The frequency spectrum and the distance dependence of  $\tilde{S}_{21f}^{yy}(\omega, r, /)$ are plotted in Fig.~\ref{AnalyticalS21versusF}  and \ref{AnalyticalS21versusR} for a illustrative parameters. The forms of these frequency- and propagation-distance-dependences can be understood by scrutinizing the different contributions within $\tilde{S}_{21f}^{yy}(\omega, r, /)$. 

The complex-valued prefactor is:
\begin{equation}
\mathcal{L}(\omega)=     \frac{\chi_\textrm{max}}{w^2}     \frac{  v_g^2 }{ 
   \left({{L_\textrm{att}} (\omega -{\omega_0})}-iv_g \right)^2} \label{prefactor}
  \end{equation}
This prefactor appears as a resonance centered at $\omega_0$ and decaying with the frequency detuning $|\omega-\omega_0|$ at a rate set by the linewidth $2 v_g/L_\textrm{att}$. Note that $\mathcal{L}(\omega)$ is symmetric with respect to $\omega_0$: the material responds also at frequencies below its uniform (i.e. $k=0$) resonance (see Fig.~\ref{AnalyticalS21versusF}). 

This prefactor multiplies a first \textit{propagating} term that includes a phase-rotation and a spatial decay operand:
\begin{equation} P_1(r, \omega)= {|L_\textrm{att}|} \times e^{i\frac{ r (\omega -{\omega_0})}{v_g}} \times e^{-\frac{r}{L_\textrm{att}}}  \label{P1}  
\end{equation}
The phase accumulation $e^{i\frac{ r (\omega -{\omega_0})}{v_g}}$ translates in clear oscillations of the real and imaginary parts of $\tilde{S}_{21f}^{yy}(\omega, r, /)$ as the frequency (hence $k$) increases, see Fig.~\ref{AnalyticalS21versusF}. They come with an exponential spatial decay. 

This operand $P_1(r, \omega)$ is multiplied by a \textit{unidirectional} term $U(r, \omega)$ that reads:
 \begin{equation} \begin{split}&U(r, \omega) = -4  \Theta_{(r)} 
 \\ &+  2 \left( \Theta_{(r-w)} e^{\frac{w}{L_\textrm{att}} + \frac{i w (\omega -{\omega_0})}{V_g}} + \Theta_{(r+w)} e^{\frac{-i w (\omega -{\omega_0})}{V_g} -\frac{w}{L_\textrm{att}}}  \right) \label{Uder} \end{split} \end{equation}
The Heaviside functions $\Theta(r)$, $\Theta(r-w)$ and $\Theta(r+w)$ indicate that energy is not sent in the direction \textit{opposite} to the group velocity $v_g$: the term $\mathcal{L}(\omega) P_1(r, \omega) U(r)$ in Eq.~\ref{S21fLinearDisp} is finite only for $r>-w$ (see Fig.~\ref{AnalyticalS21versusR}).
The three Heaviside terms in $U(r, \omega)$ indicate that the emission can be understood as performed by the center of the antenna (accounted by the term $\Theta(r)$), with correcting terms $\Theta(r\pm w)$, accounting for the extraloss/underloss $e^{\mp\frac{w}{L_\textrm{att}}}$ due to the differing distances between the position of interest $r$ and the farthest or closest antenna edges.

Finally, the expression of $\tilde{S}_{21f}^{yy}(\omega, r, /)$ includes "local terms" (free of the propagation-induced dephasing and free of the propagation-induced loss previously expressed in Eq.~\ref{P1}). They can be written from the function:
\begin{equation}L(x, \omega)= \text{sgn}{(x)} \left(+{L_\textrm{att}}-x-\frac{i x {L_\textrm{att}}(\omega -{\omega_0})}{{v_g}}\right) 
\end{equation} 
These terms are local in the sense that their sum vanishes for $|r|>w$, as illustrated in Fig.~\ref{AnalyticalS21versusR} (dotted line). These terms can be understood as the self-inductance of the emitter antenna. In practice they matter only when calculating the reflection coefficient $S_{11f}^{yy}$. Importantly, the local terms preserve the unidirectional energy flow set by the direction of the group velocity.

\subsection{Linear dispersion with negative group velocity}
If the dispersion is assumed linear but with $v_g <0$, this reverses the sign of $k_0$ and one expects $\chi_{yy}(k, /)=\chi_{yy}(-k, \backslash)$. From Eq.~\ref{S21} one can then show that:
\begin{equation}\tilde{S}_{21f}^{yy}(\omega, r, \backslash) = \tilde{S}_{21f}^{yy}(\omega, -r, /) \end{equation}
This is in line with the intuitive expectation that changing the sign of the group velocity (i.e. the direction in which energy propagates) is equivalent to exchanging the positions of the antenna emitting the power and that collecting the power. 
When $v_g<0$ the energy flow is still unidirectional but now towards the $r<w$ half space.

\section{Lemma to evaluate the response to single-sided wavevector spaces} \label{lemma}
\subsection{Splitting the full response into partial integrals}
Since the dispersion relations and/or the rf fields are generally different for positive and negative wavevectors, it is interesting to split the full transmission response of Eq.~\ref{S21} into a sum of two partial responses: 
\begin{equation} \tilde{S}_{21f}=\tilde{S}_{21+}+\tilde{S}_{21-}, \label{eq25} \end{equation}
each accounting for half of the wavevector space. Writing $I(\omega, k) = \Big[ \bar h^{*}.~[ \bar{\bar{\chi}}_{(\omega_0+v_g .k)} . \bar h] \Big] \Big( h_u^{rf}(k)\Big)^2 $ within the integrand of Eq.~\ref{S21}, we define the partial responses as $\tilde{S}_{21+}(\omega, r)  = \int_{0}^{+\infty}  I(\omega, k)  e^{-i kr} dk$ and  $\tilde{S}_{21-}(\omega, r) = \int_{-\infty}^{0} I(\omega, k)  e^{-i kr} dk$. 
They can be rewritten as:
\begin{equation}\tilde{S}_{21\pm}(\omega, r) =\int_{-\infty}^{+\infty} ~ e^{-i kr} I(\omega, k) .\Theta (\pm k)dk ,\label{trick} \end{equation}where $\Theta(k)$ is the Heaviside function.
\subsection{Lemma} 
Using the inverse Fourier transform of the Heaviside function \footnote{Maintaining the unitary character requires this time to take the classical definition (prefactor free) of the Fourier transform for $\theta(k)$}, we get: 
\begin{equation}
\left\{
\begin{array}{l}
 \tilde{S}_{21\pm}(\omega, r) = \tilde{S}_{21f}(\omega, r) \otimes \left( \frac{1}{2} \delta (r)\mp \frac{i}{ 2 \pi r}\right) \\
  ~~~~~~~~= \frac{1 }{2} \Big(\tilde S_{21f}(\omega, r) \mp i  \mathcal{HT} \left\{\tilde S_{21f}(\omega, r)\right\} \Big) \\
\end{array}
\right. \label{S21kpos}
\end{equation}
where $\delta$ is the Dirac function, $\otimes$ indicates the convolution product, and the Hilbert transform $\mathcal{HT}$ applies to the variable $r$. 

By construction, the full response $ \tilde{S}_{21f}$ is the sum of the partial responses (Eq.~\ref{eq25}). Interestingly, the lemma \ref{S21kpos} indicates that the full response can be also used to calculate the partial responses $\tilde{S}_{21+}$ and $ \tilde{S}_{21-}$. We will extensively use this trick later to combine full responses in computable cases (like for a line-shaped dispersion), in order to account for transmission responses in situations where spin waves possess non-trivial dispersion relations (like any broken-linear dispersion relation as in the $\vee$-shaped cases).
\subsection{Physical meaning of the lemma: envelope, modulation and Bedrosian theorem} 
The lemma \ref{S21kpos} indicates that if the $k<0$ branch was missing from the spectum, the spatial dependence of $\tilde{S}_{21+}(\omega, r)$ would be a space-distorted version of the full response $\tilde{S}_{21f}(\omega, r)$, the distortion being due to by the existence of the $- i \mathcal{HT}$ additional term. \\ We will see that in the situations when $\frac{\omega-\omega_0}{v_g}=k_0 \gg \{ \frac{1}{L_\textrm{att}}, \frac{1}{w} \}$, the $\tilde{S}_{21f}(r)$ spectrum contain a \textit{fast} modulation with a spatial period $2 \pi / k_0$ embedded in a \textit{slow} spatial envelope related to the spin wave decay length $L_\textrm{att}$ and the antenna efficiency function (see e.g. Eq.\ref{trivialLimit}). The real and imaginary parts of $\tilde{S}_{21f}(r)$ are in quadrature in these representative situations.

The (slow) envelope and the (fast) modulation may have almost disjoined supports in reciprocal space, such that the shape of the term $\mathcal{HT} \left\{\tilde S_{21f}(r) \right\}$ can be anticipated using the Bedrosian theorem. $\mathcal{HT} \left\{\tilde S_{21f}(r) \right\}$ will be a signal embedded in the same, almost unaltered (slow) envelope but the (fast) modulation will be translated in space by a quadrature displacement of length $ \frac{\pi}{2 k_0}$. Owing to the properties of the Hilbert transform, the sign of this displacement will depend on sgn($r$). 

The distortion induced by the $\mathcal{HT}$ terms within the partial integral changes the repartition of the signal between the two $r \in \mathbb{R}^+$ and $r \in \mathbb{R}^-$ half spaces. This is of premium importance when analyzing the non-reciprocity of spin wave transmission from one antenna to the other and vice-versa. We shall illustrate this point later with several dedicated examples.

\section{Case of V-shaped dispersion with diagonal susceptibility} \label{Vdisp}
Let us now analyse the case of SWs possessing a $\vee$-shaped dispersion relation. We still consider $p=s=0$ and an excitation by a diagonal susceptibility term only. In the relevant wavevector interval, we write $\omega=\omega_0 + v_{g+} k$ for $k>0$ and $\omega=\omega_0 + v_{g-} k$ for $k<0$ with $v_{g+}>0$ and $v_{g-}<0$. At $\omega > \omega_0$, there are now two resonant wavevectors $k_{0+} >0$ and $k_{0-}<0$.

Because of the irregular (non-differentiable) character of a $\vee$-shaped dispersion relation in $k=0$, the Eq.~\ref {chidek} describing the susceptibility is applicable only for $\omega \gtrsim \omega_0 + \Delta \omega_0$ when the spin waves are far above the gap $\omega_0$; in the gap the spin waves are evanescent. This situation will not be studied.
\subsection{General case: non-symmetric V-shaped dispersion relation} 
In the general case when $- v_{g-} \neq v_{g+}$, the full response can then be expressed as: \begin{equation}
\tilde{S}_{21f}^{yy}(\omega > \omega_0, r, \vee) = \tilde{S}_{21-}^{yy}(\backslash, v_{g-}) + \tilde{S}_{21+}^{yy}( /, v_{g+}) 
\end{equation} where each partial term is now evaluated with its own group velocity.
The partial contributions of the $k<0$ and $k>0$ branches can be calculated using the lemma \ref{S21kpos}: 
\begin{equation}
\begin{split}
\tilde{S}_{21f}^{yy}(\omega > \omega_0, \vee) = 
\frac{1 }{2} \Big(\tilde S_{21f}^{yy}(\backslash, v_{g-}) + \tilde S_{21f}^{yy}(/, v_{g+})\Big) \\
 + i \frac{1} {2} \mathcal{HT} \left\{   \tilde S_{21f}^{yy}(\backslash, v_{g-}) - \tilde S_{21f}^{yy}(/, v_{g+})    \right\}
\end{split}
\end{equation}
The second line of the previous expression cannot be made explicit in a reasonably compact form in the general case. 
\subsection{Symmetric V-shaped dispersion relation} 
However the situation simplifies when the dispersion relation is symmetric with $v_{g+} =-v_{g-}= v_g >0$ and when the susceptibility is even in $k$ with $\bar {\bar {\chi}}(k)={\bar {\bar {\chi}}}(-k)$. In this case we define $L_\textrm{att}>0$ and have: 
\begin{equation}
\begin{split}
 & \tilde{S}_{21f}^{yy}(\omega > \omega_0, r, \vee) = \\
& \frac{1 }{2} \Big(\tilde S_{21f}^{yy}(\omega, r, /, |v_g|) + \tilde S_{21f}^{yy}(\omega, -r, /, |v_g|)\Big) \\
& + i \frac{1} {2} \mathcal{HT} \left\{ \tilde S_{21f}^{yy}(\omega, -r, /, v_g)- \tilde S_{21f}^{yy}(\omega, r, /, v_g)\right\}
\end{split}
\label{veeDispS21}
\end{equation}
\subsubsection{Perfect reciprocity} 
\begin{figure}
\begin{center}
\hspace*{-0.cm}\includegraphics[width=8.5 cm]{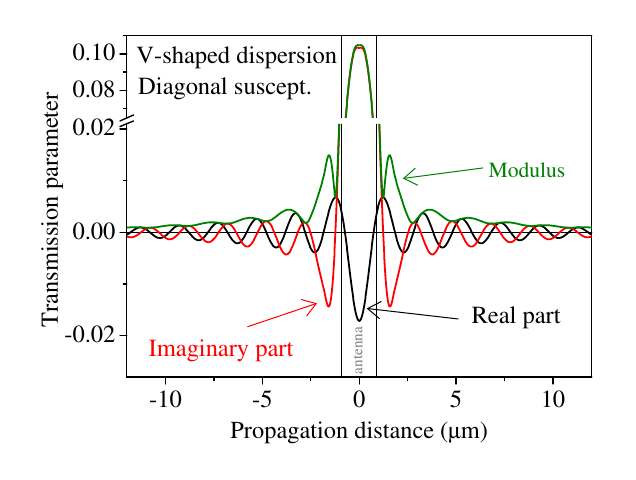}
\caption{Transmission parameter versus propagation distance in the case of a reciprocal, symmetric $\vee$-shaped dispersion relation when only one diagonal term of the susceptibility tensor is considered (Eq.~\ref{veeDispS21}). The attenuation length is 6 $\mu$m. The two resonant wavevectors are $k_0=\pm \pi~\textrm{rad/}\mu\textrm{m}$ corresponding to a spatial period of 2 $\mu$m. The antenna width is $w=1.8~\mu$m, with zero thickness $(p=0)$ and zero spacing $(s=0)$. }
\label{AnalyticalS21VversusR}
\end{center}
\end{figure}

Remembering that $\mathcal{HT}$ transforms odd functions in even functions, we can see that: 
\begin{equation} \tilde{S}_{21f}^{yy}(r, \vee) = \tilde{S}_{21f}^{yy}(-r, \vee) \end{equation}
While for a line-shaped dispersion relation, the energy was propagating in a half space only, a $\vee$-shaped dispersion relation leads to a propagation of energy in the \emph{whole} space, symmetrically with respect to the antenna center (NB: note that in this section we consider a diagonal term of the susceptibility tensor). In other words, a VNA-based electrical measurement with \textit{gedanken} antennas involving only the diagonal susceptibility terms would show a perfectly reciprocal behavior with $\tilde{S}_{21f}^{yy}(\vee)=\tilde{S}_{12f}^{yy}(\vee)$. This is illustrated in Fig.~\ref{AnalyticalS21VversusR} ; the spatial dependence of $\tilde{S}_{21f}^{yy}(\vee)$ is plotted for typical experimental parameters. 

\subsubsection{Analytics} 
Above $\omega_0 + \Delta \omega_0$, the terms of Eq.~\ref{veeDispS21} can be made explicit. The "direct" terms (line 2 of Eq.~\ref{veeDispS21}) are straightforward to understand: they create the $r$-even equivalent of Eq.~\ref{S21fLinearDisp}; i.e. the bidirectional ($r$-symetrized) version of Fig.~\ref{AnalyticalS21versusR}. They comprise propagating terms and local terms, in a manner similar to previously. These "direct" terms within $\tilde{S}_{21f}^{yy}(\omega > \omega_0, r, \vee)$ can be expressed as:
\begin{equation}
\begin{split}
&\frac{1 }{2} \Big(\tilde S_{21f}^{yy}(\omega, r, /, |v_g|) + \tilde S_{21f}^{yy}(\omega, -r, /, |v_g|)\Big) = \\
& \overbrace{\sqrt{\frac{\pi}{2}} \mathcal{L}(\omega) }^{Lorentzian} 
~~~\times \\
&\Big{[} \overbrace{P_2(r+w, \omega) + P_2(r-w, \omega)-2 P_2(r, \omega)}^{propagation~terms} + \\
&  \underbrace{\big(1+ \frac{i    {L_\textrm{att}} (\omega -{\omega_0})}{v_g}\big)\Big(  |r+w|+|r-w|-2|r| \Big)}_{local~terms} \Big{]} \\
\end{split} \label{DirectTerms}
\end{equation}
where the last line contains the local terms that depend on the positive-definite distances, and where the propagating terms depend on the function:  
\begin{equation}
P_2(x, \omega)=L_\textrm{att} e^{-|\frac{x}{L_\textrm{att}}|} \left[\Theta_{(x)} e^{\frac{-i x (\omega -{\omega_0})}{|v_g|}}  +
\Theta_{(-x)} e^{\frac{i x (\omega -{\omega_0})}{|v_g|}}  \right] \label{functionP2}
\end{equation}
The physical interpretation of Eq.~\ref{DirectTerms} is straightforward: the local terms represent some auto-inductance, and the propagating terms contain energy emitted by the center of the antenna and propagating to its both sides with the same decay length but opposite sense of phase rotation. They are complemented by other propagating terms for which the relevant travel distances are incremented or decremented by $w$ with the associated propagation loss and dephasing arising from the increment or decrement of the propagation distance. 


The $\mathcal{HT}$-transformed term (line 3 of Eq.~\ref{veeDispS21}) admits a very heavy analytical expression (see appendix, Eqs.~\ref{HTtermVshapedPrefactor}, \ref{HTtermVshapedBigFunction1}, \ref{HTtermVshapedBigFunction2}) for distant antennas (i.e. for $|r|> w$).\footnote{For $|r| < w$ (when the two antenna overlap) the line 3 of Eq.~\ref{veeDispS21} admits also an an analytical expression (not shown, but plotted in Fig.~\ref{AnalyticalS21VversusR}).}. The application of the expression is included in Fig.~\ref{AnalyticalS21VversusR}). 

\subsubsection{Dependence over propagation distance and frequency} 
From the Bedrosian theorem combined with the $\times i $ multiplication in the $\mathbb C$-plane in the line 3 of Eq.~\ref{veeDispS21}, one an anticipate that the "direct" term [line 2 of Eq.~\ref{veeDispS21} and the Hilbert-Transformed "indirect" terms [line 3] have very similar shapes, expect in the for $|r| < w$ region where the signal is dominated by the local terms (not shown). For $|r| > w$, the phase of all terms within Eq.~\ref{veeDispS21} rotate with the propagation distance and with the frequency at a pace essentially defined by the phase rotator functions $P_1(r, \omega)$ and $P_2(r, \omega)$. This is for instance clear in Fig.~\ref{AnalyticalS21VversusR}: the real and imaginary parts of the transmission signal change sign every time the propagation distance $r$ is incremented by a half period $\frac{\lambda}{2}=\frac{\pi v_g}{\omega-\omega_0}$ of the spin wave. Equivalently, they change signe every time the frequency is incremented by $v_g/(2 r)$.

Note that while a mixture of Eq.~\ref{trivialLimit}  and Eq. \ref{functionP2} is routinely\cite{ciubotaru_all_2016, talmelli_reconfigurable_2020, devolder_measuring_2021, sushruth_electrical_2020, wang_nonreciprocal_2021} used to interpret propagating spin wave spectroscopy experiments by setting simply $x=r$, we emphasize that it is not always appropriate. Such a fitting procedure indeed confuses the propagating terms  $P_2(r-w)$ and $P_2(r+w)$ with $P_2(r)$. This is valid as long as (i) one restricts to situations where $|r| \gg w$ (i.e. distant antennas) and (ii) one  can neglect the contribution of the non-diagonal terms of the susceptibility tensor. 


\section{Response to the full susceptibility tensor} \label{FullSusceptTensor} 
So far we have looked at the consequence of the sole diagonal terms of the susceptibility tensor. This is over: from now on we take into account the complexity of $\bar h$ that arises from its $z$ component that is $\propto \textrm{sgn}(k)$ (see Eq.~\ref{h}). This means that from now on, helicity mismatch is included. 
\subsection{Accounting for the helicity dependence at emission and detection} 
The factor $\Big[ \bar h^{*}.~[ \bar{\bar{\chi}}_{(\omega, k)} . \bar h] \Big]$ within the integrand of Eq.~\ref{S21} contains two kinds of terms.
\subsubsection{Regular susceptibility terms}
The diagonal terms involving $\chi_{xx, yy, zz}$ as well as $\chi_{xy, xy}$ (that happen to vanish when the applied field is along $x$ as in all the situations studied hereafter) do not depend on the sign of $k$. For all these terms, the models with the $yy$ superscript  are applicable provided the proper $\chi_\textrm{max}$ are used. The summation of Eq.~\ref{eq25} still holds: 
$$\tilde{S}_{21f}^{yy} (\omega, r) = 
\tilde S_{21+}^{yy}(\omega, r) +\tilde S_{21-}^{yy}(\omega, r)$$provided $yy$ is replaced by the considered susceptibility term among $\{xx, yy, zz,xy, xy\}$.
\subsubsection{Non-diagonal susceptibility terms involving $\bar{h}_z$ exactly once} 
In contrast, the $1z$ terms involving $\chi_{yz}$ or $\chi_{zy}$ (as well as $\chi_{xz, xz}$ that happen to vanish when the applied field is along $x$ as in all the situations studied hereafter) involve $\bar {h}_z$ exactly once; they are hence $\propto \textrm{sgn}(k)$, such that the partial integrals for the $k<0$ and $k>0$ branches must be calculated separately using the lemma \ref{S21kpos}.  For this terms the combination of $\bar {h}_z \propto \textrm{sgn}(k)$ and Eq.~\ref{eq25} results in:
\begin{equation} \tilde{S}_{21f}^{1z} (\omega, r) = 
\tilde S_{21+}^{1z}(\omega, r) - \tilde S_{21-}^{1z}(\omega, r) \end{equation}
Using the lemma~\ref{S21kpos}, we get:
\begin{equation} \tilde{S}_{21f}^{1z} (\omega, r) = - i \mathcal{HT}\{ \tilde S_{21f}^{yy} (\omega, r)\}  
\label{S21zz}
 \end{equation}where the $\chi_\textrm{max}$ in the right hand side of the equation should be replaced by either $\chi_{yz}^\textrm{max}$ or $\chi_{zy}^\textrm{max}$ depending on the term being calculated. Note that $\chi_{yz}^\textrm{max}$ and $\chi_{zy}^\textrm{max}$ are opposite and real-valued at the resonance. We will see that in many situations Eq.~\ref{S21zz} admits an analytical solution.
Two points are worth mentioning: they concern the frequency content and the spatial content of Eq.~\ref{S21zz}.

(i) The Hilbert transform applies to the space variable $r$ only. At a propagation distance $r_0$, the frequency dependence of $\tilde S_{21f}^{1z}(r_0)$ will thus \textit{always be the same} as $\tilde{S}_{21f}^{yy}(r_0)$. Because of this appearance similarity, they cannot be distinguished by a frequency scan at a single propagation distance.

(ii) The spatial content of Eq.~\ref{S21zz}: once again, if sufficiently far from the resonance, i.e. if $\left|\omega- \omega_0 \right| \gg \Delta\omega_0$, the Bedrosian theorem and the form of $\bar {\bar {\chi}}$ indicate that the spatial variations of $\tilde{S}_{21f}^{1z} (\omega, r)$ versus $\tilde{S}_{21f}^{xx,yy,zz}$ may look like scaled complex conjugates. We will show later examples illustrating this point.

\subsection{Full response in the most general case} 
The full response is the sum involving all terms of $\bar {\bar {\chi}}$: 
$$ \tilde{S}_{21f} =  \sum_{xx, yy, zz, xy, yx} \tilde{S}_{21f}^{yy} + \sum_{yz, zy, xz, zx} \tilde{S}_{21f}^{1z}$$

We remind that the static field is along the in-plane direction $x$.

The simplest case is when the total moment of the magnetic system has a longitudinal-only susceptibility (i.e. when $\bar {\bar \chi}$ reduces to a sole vanishing term which is $\chi_{xx}$). This is for instance the case for the total moment of the optical mode of a SAF in the scissor state. In this case the full response reads simply:
\begin{equation}
\tilde{S}_{21f} (\textrm{longitudinal}~\bar {\bar \chi})=  \tilde{S}_{21f}^{xx} {, ~\textrm{with}~\chi_\textrm{max} \in i\mathbb{R}^-} \label{S21allChisOPT}
\end{equation}
such that the previous methodology can be used in a straightforward manner.

A more standard situation is when both diagonal and non-diagonal susceptibility terms that are transverse to the field are involved in the dynamics. This is for instance the case for the spin-waves of in-plane magnetized single layers above saturation, or for the total moment of the acoustical mode of a SAF. In these cases we can use Eq.~\ref{S21zz} and gather the terms as follows:
\begin{equation}
\tilde{S}_{21f} (\textrm{transverse}~\bar {\bar \chi})=  \underbrace{\sum_{yy, zz} \tilde{S}_{21f}^{yy}}_{\chi_\textrm{max} \in i\mathbb{R}^-} - i  \mathcal{HT} \underbrace{\left\{\sum_{yz, zy} \tilde{S}_{21f}^{yy} \right\}}_{\chi_\textrm{max} \in \mathbb{R}} \label{S21allChisACOU}
\end{equation}

(i) The first brace involves the two susceptibility terms that are Lorentzian, with a negative imaginary value $\chi_\textrm{max}$ at the resonance. They lead to contributions $\propto S_{21f}^{yy}$, where the proportionality is a real number. These contributions resemble that of Fig.~\ref{AnalyticalS21versusFlinearDispersion} or that of Fig.~\ref{AnalyticalS21VversusR}, depending on the shape of the dispersion relation. \\

(ii) The second brace involve susceptibility terms that have a real value $\mp i \epsilon \chi_\textrm{max}$ at the resonance (see Eq.~\ref{chi}). As a result, the real parts of $\pm S_{21f}^{zy, yz}$ thus contribute to the imaginary part of the signal enclosed in curly brackets, and the imaginary parts of $\pm S_{21f}^{yz, zy}$ contribute to the real part of the signal enclosed in curly brackets. \\Beside, this signal is Hilbert-Transformed according to the space variable $r$ and then multiplied by $-i$. This process affects the spatial profile of the response, notably because the $\mathcal{HT}$ transforms even function in odd ones. However, situations may stay simple when the Bedrosian theorem is applicable, as detailed in a few examples below. 

\section{Responses including the non-diagonal parts of the susceptibility} \label{AntiDiag} 
\begin{figure}
\begin{center}
\hspace*{-0.2cm}\includegraphics[width=8.5 cm]{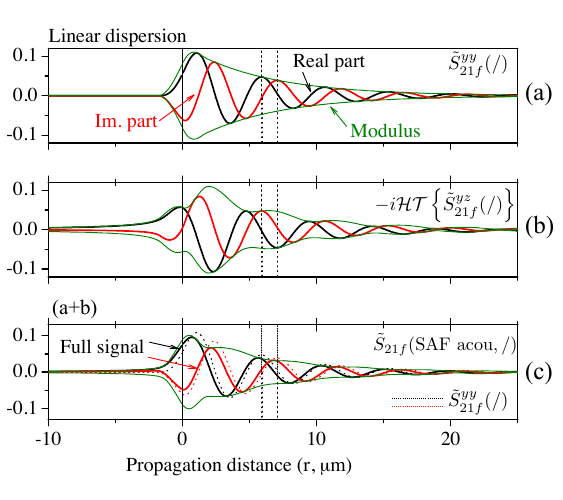}
\caption{Comparison of the two categories of terms contributing the antenna-to-antenna transmission parameter versus propagation distance for a linear dispersion relation (Eq.~\ref{S21allChisACOU}). The parameters are $p=s=0$, $w=1.8~\mu\textrm{m}$, 
$\omega_0=5$ Grad/s, $\omega=13$ Grad/s, $L_\textrm{att}=6~\mu$m, $v_g=6$ km/s. \\(a) Appearance of a diagonal term: $\tilde{S}_{21f}^{yy}(/)$ for a $\chi_\textrm{max} = -i$ i.e. corresponding to $\chi_{yy}$. 
\\(b) Appearance of a $1z$ term: $-i \mathcal{HT} \left\{ \tilde{S}_{21f}^{yy} (/) \right\} $ for a $\chi_\textrm{max} = 1$. In practice the amplitude of this term will be decremented by the ellipticity factor $\epsilon$. 
\\(c) Sum of (b) and (c) for $\epsilon / \sin(\varphi)=0.3$, accounting for full response of the acoustical spin wave branch of a SAF with a linear dispersion relation according to Eq.~\ref{S21allChisACOU}. The vertical dotted lines are guides to the eye meant to evidence the evolution of the phase of the signal upon the $-i \mathcal{HT} \left\{ \times i \right\}$ operation.}
\label{LineDispFinalPlot}
\end{center}
\end{figure}

\subsection{Line-shaped dispersion: quasi-unidirectional energy flow} 
Let's consider the case of $\tilde{S}_{21f} (\omega, r, /)$, i.e. the total response including the whole susceptibility tensor for a dispersion relation that is line-shaped in the range of wavevectors allowed by the antenna. This situation mimics the SWs within for instance a SAF in the scissors state with $
\vec k \parallel \vec H$ (see ref.~\cite{millo_unidirectionality_2023}) or within a dipolarly-coupled bilayer (see for instance Fig. 4f in ref.~\cite{qin_nanoscale_2021}). For a line-shaped dispersion relation, Eq.~\ref{S21zz} can unfortunately not be simplified and has to be integrated numerically.

An example of full signal and its contributing terms $\tilde S_{21f}^{yy}(/)$ and $\tilde S_{21f}^{1z}(/)$ is given in Fig.~\ref{LineDispFinalPlot}. The last panel compares the unidirectional case $\tilde{S}_{21f}^{yy} (\omega, r, /)$ obtained formerly for a diagonal susceptibility, with the total response including all terms of the susceptibility tensor, a real antenna and a representative ellipticity of the precession. 

While the $S_{21f}^{yy}(/)$ term was perfectly unidirectional [i.e. \textit{strictly} vanishing for $r\leq-w$, see Fig.~\ref{LineDispFinalPlot} (a)], the presence of the $1z$ terms makes the total response no longer single-sided; the $ -i\mathcal{HT}$ terms in Eq.~\ref{S21allChisACOU} add a signal that leaks into the $r \leq -w$ half space. 
The amplitude of this leak is tiny except when $\omega \approx \omega_0$ where the Bedrosian theorem is inadequate.\footnote{Indeed when $|k_0|$ is large, i.e. for frequencies obeying $\left| \omega-\omega_0 \right| \gg \Delta \omega_0 $, the Bedrosian theorem predicts that the $-i\mathcal{HT}$ terms in Eq.~\ref{S21allChisACOU} will essentially vanish for $r < -w - \frac{\pi}{2 k_0}$, as clear by  comparing Fig.~\ref{LineDispFinalPlot}(a) and (b). The ripple of the envelope in (b) is reminiscent of the $k_0 \neq 0$ oscillations in (a). This non-vanishing ripple arises because the hypotheses quantitatively required to apply the Bedrosian theorem (the \textit{disjoined} character of the spectra of the envelope and of the modulating oscillatory signal) are not strictly fulfilled.}
The unidirectional emission of the full signals $\tilde{S}_{21f}(/ \textrm{~or~} \backslash)$ is essentially maintained despite the $ -i\mathcal{HT}$ terms. This quasi-unidirectional emission, already evidenced experimentally in ref.[\cite{millo_unidirectionality_2023}], is of strong interest for applications.
\subsection{Symmetric $\vee$-shaped dispersion: amplitude non-reciprocity} 
Let us se what happens for a symmetric $\vee$-shaped dispersion relation. This represents for instance the case of the moment of a single layer film above saturation in the Damon-Eshbach geometry. This can also represent the total moment of the acoustical mode of a SAF for $k \perp H_{dc}$.
For a $\vee$-shaped dispersion relation, we have shown that: $$\tilde{S}_{21f}^{yy}(r, \vee)=\tilde{S}_{21f}^{yy}(-r, \vee).$$ This $r$-even character expresses that the $yy$ terms lead to transmission signals that are free of amplitude non-reciprocity.
However owing to the properties of the Hilbert transform, the terms below the second brace in Eq.~\ref{S21allChisACOU} will thus be $r$-odd: $$\tilde{S}_{21f}^{1z}(r, \vee)=- \tilde{S}_{21f}^{1z}(-r, \vee)$$
so that they add or subtract from contribution of the diagonal terms of the susceptibility tensor, depending on the sign of $r$. This leads to a self-evident stateme\textcolor{black}nt: the $r$-even contributions $\tilde{S}_{21f}^{yy}(r, \vee)$ and $r$-odd contributions $\tilde{S}_{21f}^{1z}(r, \vee)$ can be separated easily in an experimental measurement in the $\vee$ situation, by simply calculating $S_{21}+ {S_{12}}$ or $S_{21}- {S_{12}}$. 
Using $\{\mathcal{HT}\}^2=-1 \times$, Eq.~\ref{S21zz} and \ref{veeDispS21}, we can express $\tilde{S}_{21f}^{1z}(\omega, r, \vee)$ for $|r| \geq w$ as:
\begin{equation}
\begin{split}
\tilde{S}_{21f}^{1z} (r, \vee)  =  - \frac{i} {2}      \left( \tilde S_{21f}^{yy}(-r, /)- \tilde S_{21f}^{yy}(r, /)\right) \\
- \frac{i }{2} \mathcal{HT}\left\{  \tilde S_{21f}^{yy}(r, /) + \tilde S_{21f}^{yy}(-r, /) \right\} \\ 
\end{split}
\label{S1zV}
\end{equation}
in which on shall pay attention to use the $\chi_\textrm{max}$ of the non-diagonal susceptibility term being considered.
Only the first term can be made explicit. The second term of Eq.~\ref{S1zV} must be integrated numerically.


As a last comment, let us mention that in practice, the amplitude non-reciprocity is often measured by looking at the experimental quantity $\frac{S_{21}}{S_{12}}(r)$, which is:
\begin{equation}
\frac{\tilde S_{21f}(\omega, r)}{\tilde S_{21f}(\omega, -r)}(\vee)=\frac{{{\sum\limits_{yy~\textrm{terms}}} \tilde S_{21f}^{yy}(r) + \sum\limits_{1z~\textrm{terms}} \tilde S_{21f}^{1z}(r) }} {{{\sum\limits_{yy~\textrm{terms}}} \tilde S_{21f}^{yy}(r) - \sum\limits_{1z~\textrm{terms}} \tilde S_{21f}^{1z}(r)}} \label{VshapedANR}
\end{equation}
where we have omitted the $\vee$ labels. This corresponds to the commonly observed "amplitude non-reciprocity" of Damon-Eshbach modes \cite{schneider_phase_2008} arising from helicity match/mismatch. Note that the terminology "amplitude non-reciprocity" is somewhat improper. Indeed as soon as the non-diagonal susceptibility terms contribute to the transmission signal, the experimental quantity $\frac{S_{21}}{S_{12}}$ is then neither real nor constant: it is both position-dependent and frequency-dependent.

\section{Discussion} \label{discussion}

\subsection{Perfect Helicity mismatch and its perception in the spectral response}
An intriguing question to understand is why the concept of the angle for perfect helicity mismatch (Eq.~\ref{HmM}, or --equivalently-- the angle for no inductive detection, Eq.~\ref{nid}) seems to have been missed or ignored so far; this is intriguing since it is a rather trivial consequence of the form of the susceptibility tensor (Eq.~\ref{chi}) and the form of the rf field of an antenna (Eq.~\ref{h}).

The likely reason is that the HmM effect has no perceivable consequence in most of the situations encountered in previous works, in particular for single layer films above saturation. Indeed in the popular Damon-Eshbach configuration in single layers with or without Dzyaloshinskii interaction \cite{di_direct_2015, belmeguenai_interfacial_2015}, one uses field orientations of $\varphi=\pm \pi/2 $ far from the HmM angles, such that the HmM effect is irrelevant there. 

In contrast, the much less studied backward volume configuration of SWs in single layers corresponds to $\varphi=0\textrm{~or~}\pi$, which is much closer to $\varphi_{HmM}$. For $\varphi=0$, the backward volume SWs feature a $\Lambda$-shaped dispersion relation, with an extremely small group velocity and an almost flat dispersion relation, rendering difficult propagating spin wave spectroscopy experiments\cite{sato_propagating_2014, wessels_direct_2016, bhaskar_backward_2020}. For a single layer film, as soon as the field is slightly tilted away from the backward volume configuration (in particular when $\varphi=\varphi_{HmM}$), the SWs recover a  $\vee$-shaped dispersion relation with a forward character. When measuring $\tilde S_{21f}(r>0)$, one looks at the energy transmission from antenna 1 to antenna 2, hence essentially only due to the $k>0$ spin waves that have a group velocity in the $1 \rightarrow 2$ direction. The main (propagating) part of the signal is thus provided by the term $\frac 1 2 \tilde S_{21+}(r>0, v_g^+)$ in Eq.~\ref{veeDispS21}. For this field orientation, the HmM direction is in the $k<0$ side (or equivalently it is in the $\varphi_{HmM}+\pi$ orientation): this means that the SWs that are not excited because of the perfect helicity mismatch would anyway not send energy in the direction that one is sensitive to; the spin waves able to transfer energy between antennas are not subjected to the HmM effect. It is therefore understandable that the concept of perfect helicity mismatch has been missed in earlier studies.

\subsection{Applicability of the analytical approximations} 
A second important question is the range of applicability of the analytical approximations done after Eq.~\ref{S21}. The main approximation is the linearization of $\chi(k)$ (Eq.~\ref{chidek}) done for non-flat dispersion relations. While in the case of a line-shaped dispersion (section \ref{LinDisp}) this approximation is always legitimate, it is not near the singular point $\omega=\omega_0$ for broken-linear dispersion relations, and in particular for a $\vee$-shaped dispersion relation (section \ref{Vdisp}). This pathological change of slope of $\omega(k)$ at $k=0$ happens every time an \textit{infinitely} extended magnetic system is considered. This singularity at the dispersion relation $k=0$ is not so much of a problem in practice because real antennas are unable to generate at strictly $k=0$ (\textcolor{black}{see note [33}]). \\ 

Besides, any real magnetic system has a finite lateral size $L$ so that this singularity disappears: the dispersion relation becomes regular (differentiable) at $k=0$ and the frequency $\omega_0$ changes slightly because of the shape anisotropy associated to the finite dimension. We can thus model the $|k| < 1/L$ part of the spectrum within the flat band model, which yields would a substantial response only within the very near space the emitter antenna (see Fig.~\ref{FlatDispersionAllTerms}). \\
If in contrast one excites the system at larger $k$ i.e. "far" above\footnote{in practice with a frequency $\omega - \omega_0 \gg \{ \Delta \omega_0, v_g/L \}$} FMR, the $k=0$ singularity of the dispersion relation becomes irrelevant and we expect a response like that of Fig.~\ref{AnalyticalS21VversusR} made un-symmetrical by the distortion of Eq.~\ref{VshapedANR}. In between these two frequency domains, an intermediate behavior should be found, in which the travelling character of the spin waves get progressively perceivable in the $r$-dependence of $\tilde S_{21f}(\vee)$.

\textcolor{black}{\subsection{On the deduction of dispersion relation from transmission spectra}
Propagation Spin wave spectroscopy (PSWS) is usually conducted to get information about the dispersion relation of spin waves or at least, their group velocities. The zero-damping limit (Eq.~\ref{trivialLimit}) can be used to qualitatively interpret PSWS experiments. For a more quantitative analysis, one typically \cite{ciubotaru_all_2016, talmelli_reconfigurable_2020, wang_nonreciprocal_2021} adds a propagation loss term and use the ansatz \cite{sushruth_electrical_2020}: 
\begin{equation}  \tilde{S}_{21}^\textrm{single~mode}\big(k(\omega)\big) \propto  ~i ~e^{-i k |r|} ~e^{-\frac{|r|}{L_\textrm{att}}} ~(h_u^{rf}(k))^2. \label {1Dmodel} \end{equation}
We emphasize that it is not always appropriate. First, because generally several families of spin waves coexist in the spin wave conduit so that the phase of the transmission coefficient at a given frequency is influenced by more than one single $k$ value. This first problem can sometimes be circumvented if an appropriate group velocity-selective method is used to isolate the contribution of a family of spin waves \cite{devolder_measuring_2021}. However the use of Eq.~\ref{1Dmodel} confuses the propagating terms $P_2(r-w)$ and $P_2(r+w)$ with $P_2(r)$ in Eq.~\ref{functionP2}, which is only valid when $|r| \gg w$ (i.e. distant antennas). Besides the use of Eq.~\ref{1Dmodel} is equivalent to neglecting the contribution of the non-diagonal terms of the susceptibility tensor, such that the part of the spectrum influenced by the vicinity of $k=0$ is not modeled correctly. As a reminder, care should be taken when using Eq.~\ref{1Dmodel} to quantitatively model entire transmission spectra. } 

\textcolor{black}{As a result, the recording of solely frequency-resolved transmission spectra is generally not sufficient to deduce the dispersion relation. When trying to get the $\omega(k)$ from an experimental spectrum, one faces an additional difficulty: the phase $-i k r$ in PSWS data is known only modulo $2 \pi$. A starting point along the dispersion relation is thus needed to deduce the whole dispersion relation. One generally uses the value of $\omega(k=0)$ which --in contrast to common thinking-- cannot be extracted from the sole PSSW data. Indeed the frequency-resolved transmission spectra have qualitatively similar shapes (enveloppe and oscillations therein) for $\vee$-shaped and line-shaped dispersion relations \footnote{We can show by numerical integrations of Eq.~\ref{S21} that this kind of shape is quite general and arises as soon as the dispersion relation comprise non-flat branches} but the frequency $\omega(k=0)$ can be at various positions within the enveloppe, depending on the type of dispersion relation and the shaped of the antenna (single-wire as in Fig.1, U-shaped as in \cite{talmelli_reconfigurable_2020}, etc..). }

\textcolor{black}{For instance, the  $\omega(k=0)$ frequency is $\Delta \omega_0 \neq 0$ above the onset of the envelope for $\vee$-shaped dispersion relation (devices with a single-wire antenna harnessing Damon-Eshbach Magnetostatic spin waves \cite{ciubotaru_all_2016} or Forward volume spin waves). The  $\omega(k=0)$ frequency is $\Delta \omega_0 \neq 0$ below the roll-off of the enveloppe for devices with single-wire antenna and $\Lambda$-shaped dispersion relation (Backward volume spin waves). It is the middle of the enveloppe for devices with single-wire antenna and line-shaped dispersion relation. (Fig.~\ref{AnalyticalS21versusFlinearDispersion}). In all the other cases (more complex form of dispersion relation and/or antenna emitting a spectrum that is not maximal at $k=0$), the $\omega(k=0)$ frequency can be anywhere within (or without) the envelope and therefore cannot be identified from the sole knowledge of a PSWS spectrum. For good practice, $\omega(k=0)$ frequency ought to be measured separately, optimally by classical FMR.}

\section{One-way filters that are tunable between low-pass, all-pass and high-pass} 
\label{Applications}
\begin{figure}
\begin{center}
\hspace*{-0.5cm}\includegraphics[width=8 cm]{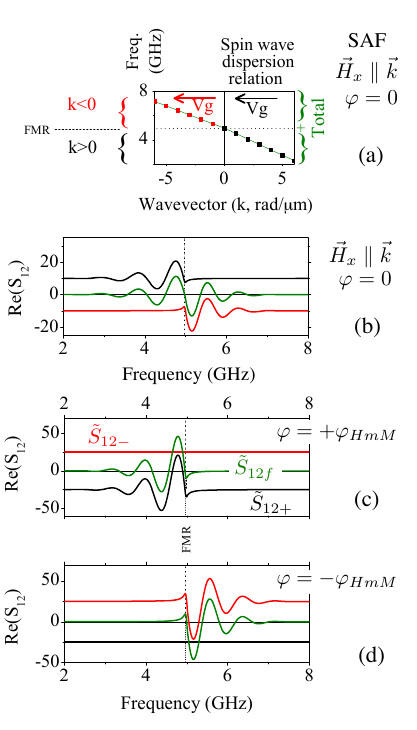}
\caption{Illustration of the combination of single-sided $k$ emission and unidirectional energy flow. (a) Linearized dispersion relation of the acoustical mode of a synthetic antiferromagnet for a wavevector parallel to the applied field ($\varphi=0$), according to ref.~\cite{millo_unidirectionality_2023}. The material parameters are from \cite{mouhoub_exchange_2023} with $\alpha=0.001$ and $\mu_0 H_x$=50 mT. (b-d): Numerical evaluations (Eq.~\ref{S21}) of the partial responses $S_{12+}$, $S_{12-}$ and the full response $S_{12f}$ for an antenna geometry with $r=3~\mu$m, $w$=0.8 $\mu$m, $p$=$s$=0 . In (b), the dc field is oriented at $\varphi=0$, such that only $h_z^{rf}$ contributes. In (c) the dc field is tilted by -16 degrees to achieve perfect helicity mismatch for the $k<0$ branches, thereby achieving a "high-pass" behavior. The field rotation is opposite in panel (d). The dispersion relations for (c) and (d) are qualitatively similar to (a) and they are taken from ref.[\onlinecite{millo_unidirectionality_2023}].}
\label{PartialKsFig}
\end{center}
\end{figure}

\textcolor{black}{When using spin waves for signal or information processing in nanodevices, the main engineering issue is the overall energy efficiency. Solutions have been developed to focus the spin wave energy \cite{vlaminck_spin_2023, kiechle_spin-wave_2023} at a particular position. Solutions have also been found to compensate for the propagation loss of spin wave energy \cite{merbouche_true_2023}. However, the bottleneck remains the low power efficiency at transduction back and forth from the electrical domain to the spin wave domain: a gain between one and two orders of magnitude is required. Ensuring a unidirectional energy flow of the spin wave energy is a step in the right direction because this avoids wasting half if the energy that the spin waves would be otherwise radiating in the unwanted direction.}
\subsection{Combining single-sided $k$ emission with unidirectional energy flow} 
Besides, the unidirectional energy flow can be combined with the single-sided $k$ emission to design one-way reconfigurable frequency filters. Let us illustrate this assuming the quasi-line-shaped dispersion relation of a SAF with $\vec k \parallel \vec H_{x, dc}$, which offers $\textrm{sgn}(v_g^+) \approx \textrm{sgn}(v_g^-)$, the two being for instance negative to ensure a unidirectional energy flow from antenna 2 to antenna 1 [see the example of Fig.~\ref{PartialKsFig}, (a)]. Indeed in this situation, the transmission $\tilde S_{21f}(r>0)$ above $\omega_0$ is due to the $k<0$ branch, while the transmission below $\omega_0$ is due to the $k>0$ branch. The sum of the two partial contributions lead to a transmission spectrum that extends symmetrically around the uniform resonance $\omega_0$, as was the case in Fig.~\ref{AnalyticalS21versusFlinearDispersion}. We can name this behavior "all-pass", and it can be obtained for the total moment of the acoustical SW mode of a SAF with $\vec k \parallel \vec H_x$, i.e. $\varphi=0$ as demonstrated experimentally in ref.~[\onlinecite{millo_unidirectionality_2023}].  

For the acoustical mode of a SAF, slightly turning the applied field from $\varphi=0$ to $\varphi=\varphi_{HmM}$ maintains\cite{millo_unidirectionality_2023} the property $\textrm{sgn}(v_g^+)= \textrm{sgn}(v_g^-)$: the dispersion relation is broken-linear but still monotonous across $k$=0. Since $v_g^+ \neq v_g^-$, none of the analytical limits described above applies, and a numerical integration of Eq.\ref{S21} must be performed. For this new field orientation $\varphi_{HmM}$, the perfect helicity mismatch cancels the contribution from the $k <0$ branch, such that the transmission is almost entirely suppressed above $\omega_0$. This "low-pass" filtering can be changed to "high-pass" filtering by rotating the dc field in the opposite direction. Examples of this "all-pass", "low-pass" and "high-pass" behaviors are shown in Fig.~\ref{PartialKsFig}. Note that the frequency filtering is obtained while maintaining a unidirectional energy transfer, similar to that of Fig.~\ref{LineDispFinalPlot}.

\subsection{Alternative materials for reconfigurable filtering} 
\textcolor{black}{The previous proposition --one-way filters that are tunable between low-pass, all-pass and high-pass-- relies on two ingredients: a monotonous dispersion relation so that the frequencies of $k<0$ spin waves and $k>0$ spin waves differ, and the ellipticity of the inductive response of the total moment of the magnetic system. While the second ingredient is fairly easy to obtain, the first one has so far been demonstrated only at low $k$'s in bilayers~\cite{qin_nanoscale_2021} and in SAFs\cite{millo_unidirectionality_2023}. Since it relies on the layer-to-layer dipolar interaction, monotonous dispersion relations should occur in many more magnetic systems that can be prepared in states that are analogs of the antiparallel state or of a scissors state. This task of searching for systems that can host unidirectional spin waves exceeds the scope of our present study. However, it would be interesting to investigate in particular several systems. This includes for instance bilayers in which the two layers are exchanged-biased in opposite directions. This includes also antiferromagnetically coupled multilayers ("multiSAFs"), as well as simple multilayers that would comprise layers that possess a contrast of coercivity/anisotropy so that the multilayer can be prepared in non-trivial magnetic configurations. }

\section{Conclusions}\textcolor{black}
This paper extends the formalism of spectroscopy of propagating spin waves when using inductive antennas connected to a vector network analyzer. The problem is supposed to be invariant in the plane transverse to the spin wave wavevector. 

The paper first revisits an apparently simple problem: the magnetic field generated by a single-wire antenna and how it excites the magnetization precession. From a reciprocal space analysis of the antenna field, we describe the phenomenon of helicity match/mismatch between the dynamic magnetization of a spin wave and the magnetic field radiated by the antenna. From the form of the susceptibility tensor reflecting the precession ellipticity, we identify specific orientations of the wavevector for which a perfect helicity mismatch is reached, such that the antenna cannot excite this spin wave, while it can for the opposite wavevector -- this single-sided wavevector generation is not to be confused with a unidirectional emission of energy. \\ 

We then analyze the stray field emanating from a spin wave, and how this stray field couples to an antenna used as an inductive detector. It appears that spin waves of properly chosen helicity do not emit any stray field on one side of the film. Inductive detection of these spin waves is impossible. These spin waves are also the ones leading to perfect helicity mismatch. In simple words, if an antenna cannot excite a given spin-wave, it cannot detect it inductively: they do not couple and are dark to each other. \\

In the second part of the paper, we combine emission, propagation and detection of spin waves to issue a formalism to calculate of the antenna-to-antenna transmission. This formalism is applied to situations that lead to physically-transparent analytical expressions. These expressions clarify the distinct roles of the direction of the wavevector of the spin waves, and of the direction of their group velocity. \\
Three canonical situations are examined: when spin waves have a flat dispersion relation, a $\vee$-shaped dispersion relation and finally a line-shaped dispersion relation. In the first case (vanishing group velocity), the spin waves  radiate energy in a slightly non-reciprocal manner but in an only evanescent way. In contrast, for the very often encountered situation of a $\vee$-shaped dispersion relation, the emission of spin waves is bidirectional and comes with some non-reciprocity. This non-reciprocity is mainly of amplitude type, but varies with frequency and propagation distance.

For spin waves with a line-shaped dispersion relation in the range of wavevectors allowed by the antenna geometry, the spin waves transport power in a quasi-unidirectional manner. This happens for the acoustical spin waves of synthetic antiferromagnets when the spin wave wavevector is close to parallel to the applied field. This situation is of interest for applications, and can be used to design filters that can be reconfigured to be low-pass, high-pass or all-pass. The present formalism offers a simple and direct method to understand, design and optimize devices harnessing propagating spin waves, including when a unidirectional energy flow is desired, and when a frequency-filtering behavior is targeted. 

I acknowledge the French National Research Agency (ANR) under Contract No. ANR-20-CE24-0025 (MAXSAW). I thank the authors of ref.~\cite{sushruth_electrical_2020} for discussion during the elaboration of earlier generations of the present model, and Claude Chappert for insightful comments.


\section*{Appendix} 
\subsection{Field radiated by a antenna of rectangular cross section}
The field radiated by a rectangular antenna of finite thickness $p$, finite width $w$ and placed at a spacing $s$ from the film has two components (Fig.~\ref{AntennaGeometry}). The in-plane component $H_u^{rf}$ is the sum of two terms. 
Defining $\Sigma=\frac{1}{2 \pi} \frac{\mu_0 I^{rf}} {w}$ as scaling factor that has the dimension of a flux density, the first term reads: \\
\begin{widetext}
\begin{equation}
\Sigma \left( \frac{p+z}{z} \left[\tan ^{-1}\left(\frac{\frac{w}{2}+u}{p+z}\right)-\tan^{-1}\left(\frac{u-\frac{w}{2}}{p+z}\right) \right] +  \frac{z}{p}   \left[\tan^{-1}
\left(\frac{u-\frac{w}{2}}{z}\right) - \tan^{-1}\left(\frac{\frac{w}{2}+u}{z}\right)\right] \right) \label{HxdirectSpace1}
\end{equation} 
The second term of $H_u^{rf}$ is:
\begin{equation}
\Sigma\left[
\frac{\frac{w}{2}-u}{2p} \ln \left(\frac{p^2+2 p z}{\left(u-\frac{w}{2}\right)^2+z^2}+1\right)+\frac{\frac{w}{2}+u}{2p} \ln \left(\frac{p^2+2 p z}{\left(\frac{w}{2}+u\right)^2+z^2}+1\right) \right] \label {HxdirectSpace2}
\end{equation}
The out-of-plane component $H_z^{rf}$ is also the sum of two terms. The first is:
\begin{equation} \Sigma \left( \frac{\left(u-\frac{w}{2}\right)}{p} \left[\tan^{-1}\left(\frac{p+z}{u-\frac{w}{2}}\right)-\tan^{-1}\left(\frac{z}{u-\frac{w}{2}}\right)\right]-\frac{\left(\frac{w}{2}+u\right)}{p}    \left[\tan ^{-1}\left(\frac{p+z}{\frac{w}{2}+u}\right)-\tan^{-1}\left(\frac{z}{\frac{w}{2}+u}\right)\right] \right) \end{equation} \label{HzdirectSpace1}
While the second term of $H_z^{rf}$ is:     
\begin{equation} \Sigma \left( \frac{p+z}{2h} \ln \frac{\frac{\left(u-\frac{w}{2}\right)^2}{(p+z)^2}+1}{\frac{\left(\frac{w}{2}+u\right)^2}{(p+z)^2}+1} +\frac{z}{2p}  \ln \frac{\frac{\left(\frac{w}{2}+u\right)^2}{z^2}+1}{\frac{\left(u-\frac{w}{2}\right)^2}{z^2}+1} \right)  \label{HzdirectSpace2} \end{equation}
\end{widetext}
\subsection{Flat horizontal dispersion with diagonal susceptibility} \label{FlatDisp}
Let us consider the hypothetical case when the SWs would have a perfectly flat dispersion relation (i.e. with $\omega(k)=\omega_0, \forall k$) such that no resonant $k_0$ can be defined. 
In this section, we assume a realistic antenna with a finite thickness (i.e. $p \neq 0$) and at a finite spacing from the film (i.e. $s \neq 0$). In the subsections 1 and 2, we start by considering that the magnet is excited by only a diagonal susceptibility term (this assumption will be removed in subsection 3).
Under these assumptions, the frequency and spatial dependences of $\tilde{S}_{21f}^{yy}(\textrm{--})$ become separable:
\begin{equation}
\begin{split}
\tilde{S}_{21f}^{yy}(\omega, r, ~ \textrm{--}) \propto & \Big[ \bar h^{*}.~[ \bar{\bar{\chi}}{(\omega_0)} . \bar h] \Big] \int_{-\infty}^{\infty} dk e^{-i kr}~  
\Big( h_u^{rf}(k)\Big)^2 \\  
&=  \tilde{s}_{21}^{yy}(\omega_0) \times s_{21}^{yy}(r,~ \textrm{--}) 
\end{split}
\end{equation}
\subsubsection{Frequency dependence for a flat dispersion and a diagonal susceptibility} 
The complex-valued frequency dependence $\tilde{s}_{21}^{yy}(\omega_0)$ is simply that of the considered diagonal element of $\bar {\bar \chi}(\omega, k=0)$, i.e. a Lorentzian response centered at $\omega_0$, exactly like in an FMR measurement. 
\subsubsection{Spatial dependence for a flat dispersion and a diagonal susceptibility} 
\begin{figure}
\begin{center}
\hspace*{-0.2cm}\includegraphics[width=8 cm]{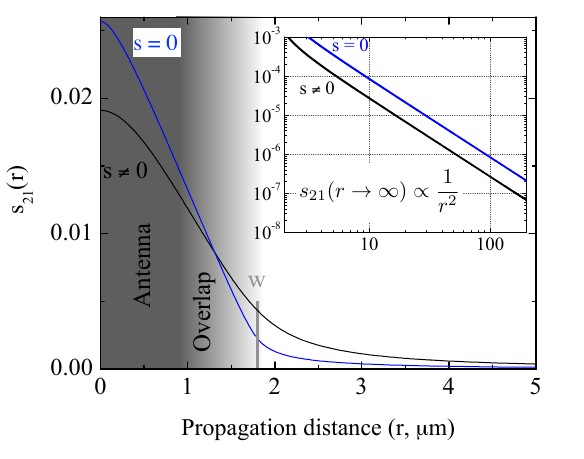}
\caption{Spatial decay $s_{12}^{yy}(r, ~\textrm{--})$ for a flat dispersion relation and a response to the diagonal susceptibility terms only, according to Eq.~\ref{decay}. Inset: idem for large propagation distances. The antenna has a width $w=1.8~\mu$m and a thickness $p=0.15~\mu$m.} 
\label{DecayPlot}
\end{center}
\end{figure}
The distance dependence $s_{21}^{yy}(r)$ is a real-valued, even, positive function that depends only on the antenna geometry.
It is the convolution of the inverse Fourier transforms 
of each of the three factors of the squared Eq.~\ref{AntennaField}: \\(i) the inverse Fourier transform of the sinc$^2$ factor, which is $\frac{\sqrt{2\pi}}{ w} \Lambda(\frac{r}{w})$, where $\Lambda(r)$ is the unit triangular function vanishing at $|r| \geq w$, \\(ii) the inverse Fourier transform of $e^{-2 |k|s}$, i.e. $\frac{4}{\sqrt{2\pi }} \left(\frac{ s}{r^2+4s ^2} \right)$, and \\(iii) the inverse Fourier transform of the finite-antenna-thickness factor $\left({1-e^{-p \left| k\right|}}    \right) /   \left(    p \left| k\right| \right)^2$, which is can be expressed from the function $g$ defined as: \\
\begin{equation} \begin{split} &{\sqrt{2 \pi }}  . g(\frac p r)= 2 r \left(\tan ^{-1}\left(\frac{2 p}{r}\right)+2 \tan ^{-1}\left(\frac{r}{p}\right)-\pi
   \right)+ \\ &4 p \left(\coth ^{-1}\left(\frac{3 p}{p-2 i r}\right)+\coth    ^{-1}\left(\frac{3 p}{p+2 i r}\right)\right)  \end{split} \end{equation} 
Overall, the spatial decay of the spin wave signal is:
\begin{equation}
s_{21}^{yy}(r, \textrm{--}) = (2 \pi)^{-\frac 3 2} \underbrace{\left[\frac{4}{ w}. \Lambda(\frac{r}{w}) \right]}_{w-\textrm{dependent}} \otimes \underbrace{\left(\frac{ s}{r^2+2s^2} \right)}_{s-\textrm{dependent}} \otimes \underbrace{g(\frac p r)}_{p-\textrm{dep.}}
\label{decay}
\end{equation}
A (very heavy) analytical formulation of this spatial decay exists. However the notation of Eq.~\ref{decay} is much more insightful since its three factors depend on different geometrical terms: the antenna width $w$, its spacing with the film $s$ and the antenna thickness $p$. \\

In case of (very unpractical experimentally !) partial overlap between the emitting and the receiving antenna, i.e. for $|r| \leq w$, the spatial decay ressembles $ \Lambda(r/w)$, hence linearly decreasing with a zero intercept for $|r| =w$ (see Fig.\ref{DecayPlot}). \\
The convolution with the Lorentzian function of width $s$, and the cusp-shaped function $g$ that has a width related to the antenna thickness $p$, makes the function $s_{21}^{yy}(r, \textrm{--}) $ not vanishing for $|r| \gg w$ (see Fig.~\ref{DecayPlot}, inset).
If $|r| \gg w$ and $p=0$ then one can show that $$s_{21}^{yy}(r \rightarrow \pm \infty, \textrm{--}) \propto \frac{1}{r^2}$$
This scaling can be understood as the consequence of the $1/r^2$ decay of the antenna in-plane field, as already noticed in ref.~\cite{sushruth_electrical_2020}. This long-range exciting field (directly) induces (unretarded) magnetization dynamics below the receiving antenna even if SWs originating from the emitter antenna never travel there because of their vanishing ability to transport energy (i.e. $v_g=0$). \textcolor{black}{Note that other mechanisms  --like antenna-to-antenna direct inductive coupling inducing current at the receiving antenna-- also contribute to the distant excitation of quasi-uniform spin waves~\cite{greil_secondary_2023}. This second mechanism is ignored here but for both these scenarios the mutual inductance of the two antennas includes a contribution from the sample susceptibility at $k=0$ that is detectable experimentally~\cite{devolder_measuring_2021}.}

\subsubsection{Spatial decay for a flat dispersion relation and all susceptibility terms } 
\begin{figure}
\begin{center}
\hspace*{-0.5cm}\includegraphics[width=9 cm]{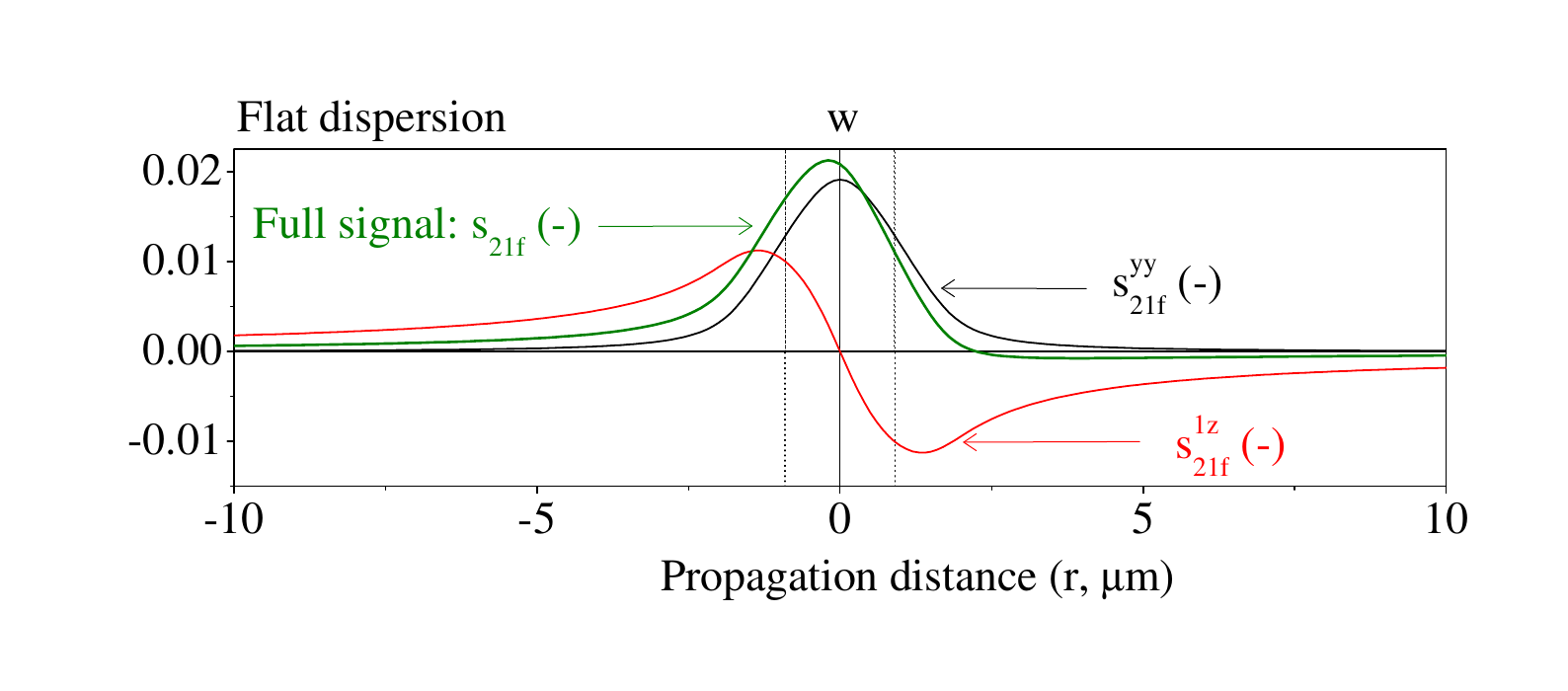}
\caption{Antenna-to-antenna transmission parameter versus propagation distance for a flat dispersion relation: the real-valued term $s_{21}^{yy}$ (Eq.~\ref{decay}) normalized by $\chi_{max}=-i$, the real-valued term $-i \mathcal{HT}\left \{s_{21}^{1z}\right\}$ normalized by $\chi_{max}=1$ and the full spatial decay $\tilde{S}_{21f}(\omega, r, \textrm{--}) / s_{21f}(\omega, r, \textrm{--})$ when $\epsilon/ \sin(\varphi)=0.3$. The parameters are $p=s=0.15~\mu$m and $w=1.8~\mu\textrm{m}$. }
\label{FlatDispersionAllTerms}
\end{center}
\end{figure}

Finally, let us include the $1z$ terms in the full response when the dispersion relation is assumed flat. It reads: 
\begin{equation}
\begin{split}
&S_{21f}(\omega, r, \textrm{--})= \tilde{s}_{21}(\omega) \times \\ 
&\left(
\underbrace{\sum_{xx, yy, zz} s_{21}(r, \textrm{--})}_{\chi_\textrm{max} \in i\mathbb{R}^-} 
 -i \mathcal{HT} \underbrace{\left \{ \sum_{yz, zy} s_{21}(r, \textrm{--})\right\}}_{\chi_\textrm{max} \in \mathbb{R}} 
\right)
\end{split} \label{flat}
\end{equation}
where the proper $\chi_\textrm{max}$ terms should be put on each term of the sums.
Fig.~\ref{FlatDispersionAllTerms} illustrates how the two terms of Eq.~\ref{flat} depend on the propagation distance. Looking back at Fig. \ref{AntennaFieldsPlot}(a), we see that the quantities $s_{21f}^{yy}(\textrm{--})$ and $s_{21f}^{1z}(\textrm{--})$ look as a smoothed versions of the real space profile of the antenna fields $h_u^{rf}$ and $h_z^{rf}$. The smoothing is essentially a convolution performed using the largest quantity among the antenna width, thickness and spacing (see Eq.~\ref{decay}).  

The frequency content of $\tilde S_{21f}(\textrm{--})$ is still a Lorentzian function centered at $\omega_0$. However the spatial content (Fig.~\ref{FlatDispersionAllTerms}) is no longer $r$-symmetric, since the $\mathcal{HT}$ terms are $r$-odd. Somehow, even if the energy of the SWs leaks in an only-evanescent way to the left and the right of the antenna, the helicity match/mismatch is still resulting in a slightly stronger transmission $\tilde S_{21f}(r, \textrm{--})$ towards the side of best helicity matching. 

\begin{widetext}
\subsection{Indirect terms for a V-shaped dispersion relation and a diagonal susceptibility term} 
Writing $E_1(\upsilon)$ the exponential integral function of the complex numbers 1 and $\upsilon$, and G is the Meijer G function, the "indirect" term $ + i \frac{1} {2} \mathcal{HT} \left\{ \tilde S_{21f}^{yy}(\omega, -r, /, v_g)- \tilde S_{21f}^{yy}(\omega, r, /, v_g)\right\} $ (i.e. line 3 of Eq.~\ref{veeDispS21}) is equal to the following prefactor:
\begin{equation}
\frac{i}{\sqrt{2 \pi }}  L_\textrm{att} \mathcal{L}(\omega) \times  \exp \left(-\frac{2 w+x}{L_\textrm{att}}-\frac{i (w (5 \omega +\omega_0)+x (\omega +\text{$\omega
   $0}))}{V_g}\right) \times (e^{\frac{w}{L_\textrm{att}}})
\label{HTtermVshapedPrefactor}  
 \end{equation}
multiplied by the sum of two terms, which are: 
First term: \\
\begin{equation}
\label{HTtermVshapedBigFunction1}
\begin{split}   
& -2 e^{w \left(\frac{1}{L_\textrm{att}}+\frac{i (5 \omega +\omega_0)}{V_g}\right)}
   \left(e^{\frac{2 i x \omega_0}{V_g}} E_1\left(-\frac{x (V_g+i L_\textrm{att} (\omega -\omega_0))}{L_\textrm{att} V_g}\right)+e^{2 x
   \left(\frac{1}{L_\textrm{att}}+\frac{i \omega }{V_g}\right)} E_1\left(x \left(\frac{i (\omega -\text{$\omega
   $0})}{V_g}+\frac{1}{L_\textrm{att}}\right)\right)
   \right) \\
&   -2 e^{w \left(\frac{1}{L_\textrm{att}}+\frac{i (5 \omega +\omega_0)}{V_g}\right)}
   \left(
   -\ln \left(1-\frac{w^2}{x^2}\right) e^{x \left(\frac{1}{L_\textrm{att}}+\frac{i (\omega +\text{$\omega
   $0})}{V_g}\right)}\right) \\
&   +e^{\frac{2 x}{L_\textrm{att}}+\frac{2 i (w (2 \omega +\omega_0)+x \omega )}{V_g}} E_1\left(-\frac{(w-x) (V_g+i
   L_\textrm{att} (\omega -\omega_0))}{L_\textrm{att} V_g}\right) \\
&   -e^{\frac{2 i x \omega_0}{V_g}} \left(e^{\frac{2 i w (2 \omega +\text{$\omega
   $0})}{V_g}}-2 e^{w \left(\frac{1}{L_\textrm{att}}+\frac{i (5 \omega +\omega_0)}{V_g}\right)}\right) E_1\left(\frac{(w-x) (V_g+i L_\textrm{att}
   (\omega -\omega_0))}{L_\textrm{att} V_g}\right) \\
&   +e^{\frac{2 i (\omega_0 (w+x)+2 w \omega )}{V_g}} E_1\left(-\frac{(w+x) (V_g+i
   L_\textrm{att} (\omega -\omega_0))}{L_\textrm{att} V_g}\right) \\
&   +e^{\frac{2 (w+x)}{L_\textrm{att}}    +\frac{2 i \omega  (3 w+x)}{V_g}} E_1\left(\frac{(w+x)
   (V_g+i L_\textrm{att} (\omega -\omega_0))}{L_\textrm{att} V_g}\right)
   \end{split}
\end{equation}
and second term: \\
\begin{equation}
\begin{split}   
   &-\pi  (e^{-\frac{w}{L_\textrm{att}}})
   \left(e^{w \left(\frac{1}{L_\textrm{att}}+\frac{i \omega
   }{V_g}\right)}-e^{\frac{i w \omega_0}{V_g}}\right)^2 e^{\frac{x}{L_\textrm{att}}+\frac{i (w (3 \omega +\omega_0)+x (\omega +\text{$\omega
   $0}))}{V_g}} G_{2,3}^{2,1}\left(-\frac{(w-x) (V_g+i L_\textrm{att} (\omega -\omega_0))}{L_\textrm{att} V_g}|
\begin{array}{c}
 0,-\frac{1}{2} \\
 0,0,-\frac{1}{2} \\
\end{array}
\right)
\end{split}
\label{HTtermVshapedBigFunction2}
\end{equation}

\end{widetext}

%

\end{document}